\documentclass[10pt,twocolumn,letterpaper]{article}

\usepackage{iccv}              

\usepackage{makecell}
\usepackage{bbm}
\usepackage{cuted}
\usepackage{graphicx}
\usepackage{amsmath}
\usepackage{amssymb}
\usepackage{booktabs}
\usepackage{newfloat}
\usepackage{listings}
\usepackage{subcaption}
\usepackage{cuted}
\usepackage{colortbl}
\usepackage{multirow}
\usepackage{booktabs}
\usepackage{mathtools}
\usepackage{bm, amsthm, amssymb}
\usepackage{filecontents}
\usepackage{lipsum}
\usepackage{etoolbox}
\usepackage{amsmath,amsfonts}
\usepackage{algorithmic}
\usepackage{algorithm}
\usepackage{array}
\usepackage{textcomp}
\usepackage{stfloats}
\usepackage{url}
\usepackage{verbatim}
\usepackage{graphicx}
\usepackage{bm, amsthm, amssymb}
\usepackage[symbol]{footmisc}
\usepackage{threeparttable}

\definecolor{iccvblue}{rgb}{0.21,0.49,0.74}
\usepackage[pagebackref,breaklinks,colorlinks,allcolors=iccvblue]{hyperref}

\def\paperID{7490} 
\def\confName{ICCV}
\def\confYear{2025}

\title{GeoAvatar: Adaptive Geometrical Gaussian Splatting for 3D Head Avatar}

\author{SeungJun Moon$^{1}$\footnotemark[1], Hah Min Lew$^{1}$\footnotemark[1], Seungeun Lee$^{1}$, Ji-Su Kang$^{1}$, and Gyeong-Moon Park$^{2}$\footnotemark[2]\\
$^1$Klleon AI Research, $^2$Korea University\\
{\tt\small $^{1}$\{seungjun.moon, hahmin.lew, seungeun.lee, jisu.kang\}@klleon.io} {\tt\small $^{2}$gm-park@korea.ac.kr}
}

\newcommand{\stdv}[1]{\scriptsize$\pm$#1}

\begin{document}
\maketitle

\begin{strip}
\centering
\vspace{-1.5cm}
\includegraphics[width=1\textwidth]{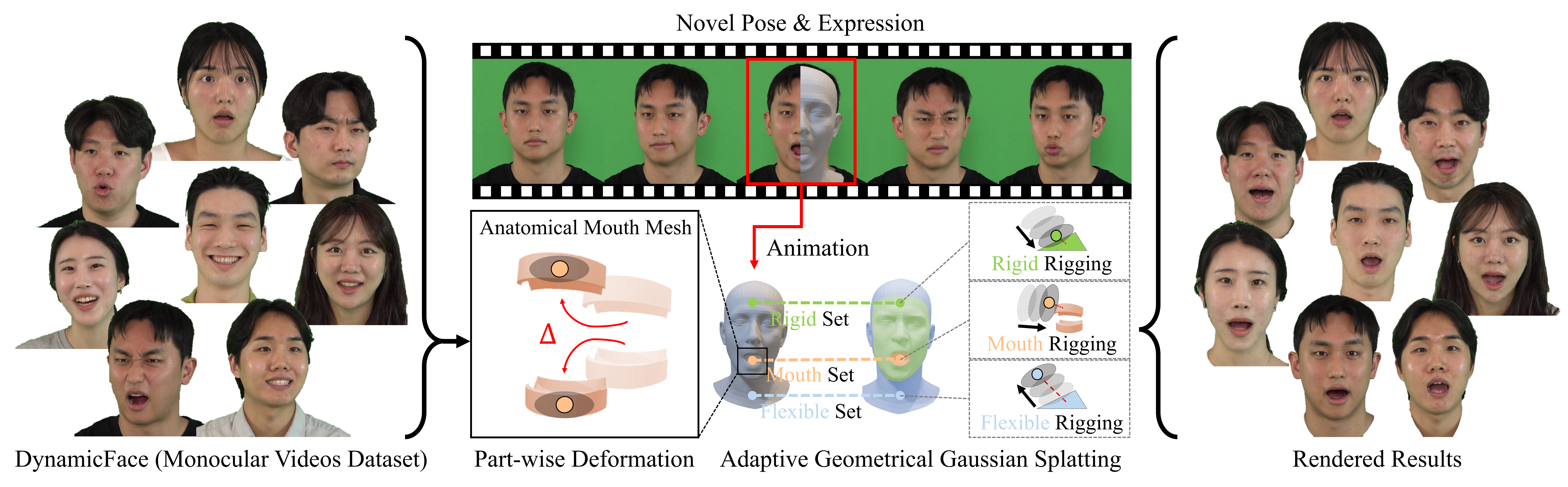}
\vspace{-0.5cm}
\captionof{figure}{\textbf{Overview of GeoAvatar.} We propose GeoAvatar, a novel adaptive geometrical Gaussian Splatting framework, and release a new monocular video dataset, DynamicFace. Given an input novel animation, GeoAvatar generates robust and high textured 3D head avatars.
}
\label{fig:front}
\vspace{-0.3cm}
\end{strip}

\begin{abstract}
Despite recent progress in 3D head avatar generation, balancing identity preservation, \ie, reconstruction, with novel poses and expressions, \ie, animation, remains a challenge.
Existing methods struggle to adapt Gaussians to varying geometrical deviations across facial regions, resulting in suboptimal quality.
To address this, we propose GeoAvatar, a framework for adaptive geometrical Gaussian Splatting.
GeoAvatar leverages Adaptive Pre-allocation Stage (APS), an unsupervised method that segments Gaussians into rigid and flexible sets for adaptive offset regularization.
Then, based on mouth anatomy and dynamics, we introduce a novel mouth structure and the part-wise deformation strategy to enhance the animation fidelity of the mouth.
Finally, we propose a regularization loss for precise rigging between Gaussians and 3DMM faces.
Moreover, we release DynamicFace, a video dataset with highly expressive facial motions.
Extensive experiments show the superiority of GeoAvatar compared to state-of-the-art methods in reconstruction and novel animation scenarios.

\end{abstract}

{\renewcommand*{\thefootnote}{}%
  \footnotetext{Project page: \href{https://hahminlew.github.io/geoavatar/}{https://hahminlew.github.io/geoavatar}}
{\renewcommand*{\thefootnote}{\fnsymbol{footnote}}\stepcounter{footnote}%
  \footnotetext{Equal contribution}
{\renewcommand*{\thefootnote}{\fnsymbol{footnote}}\stepcounter{footnote}%
  \footnotetext{Corresponding author}
\setcounter{footnote}{0}

\section{Introduction}
Recent advancements in deep learning-based 3D modeling techniques \cite{mildenhall2021nerf, kerbl20233d, muller2022instant, huang20242d} have spurred active research into their applications for 3D head avatar generation \cite{zielonka2023instant, qian2024gaussianavatars, shao2024splattingavatar, xiang2023flashavatar, sun2023next3d, zheng2023pointavatar, xuan2024faghead, chan2022efficient, xu2024gaussian, wang2024mega, ma20243d, zhou2024headstudio, grassal2022neural, li2024generalizable, zheng2022avatar, gafni2021dynamic, giebenhain2023learning, wang2023gaussianhead, dhamo2025headgas, zheng2024headgap, li2024talkinggaussian, cho2024gaussiantalker, zhao2024psavatar, chen2024monogaussianavatar, yuan2024gavatar, xu2023omniavatar, kirschstein2023nersemble, bao2024geneavatar, chai2023hiface, dib2021towards, xu2023avatarmav}.
Among these, Gaussian Splatting~\cite{kerbl20233d}, originally developed for reconstructing static scenes, has been used due to its explicit representations of Gaussians in 3D space.

Since adapting the reconstruction-oriented approach for animation poses significant challenges, recent methods have leveraged the deformable 3D Morphable Model (3DMM) \cite{loper2023smpl, li2017learning} as a 3D prior~\cite{qian2024gaussianavatars, shao2024splattingavatar, xiang2023flashavatar, zheng2024headgap, dhamo2025headgas} to extend its applicability to head avatar animation.
However, we emphasize that a critical challenge still lies in balancing high-fidelity identity preservation, \ie, reconstruction, with the robustness of various novel poses and expressions, \ie, animation.

\begin{figure*}[t]
\begin{center}
\begin{subfigure}{0.8\columnwidth}
\includegraphics[width=\columnwidth]{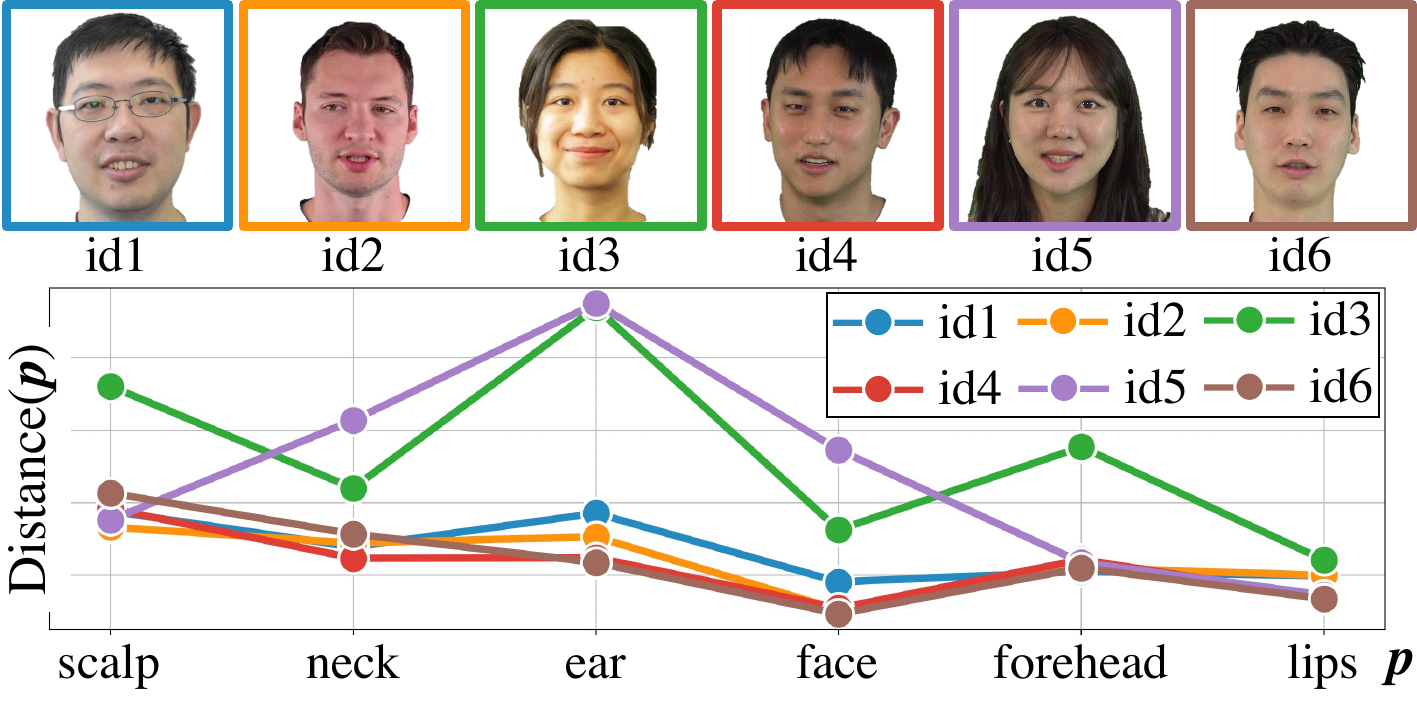}
\caption{Local mean distributions of Gaussians per facial region.}
\label{fig:motivation_a}
\end{subfigure}
\begin{subfigure}{0.8\columnwidth}
\includegraphics[width=\columnwidth]{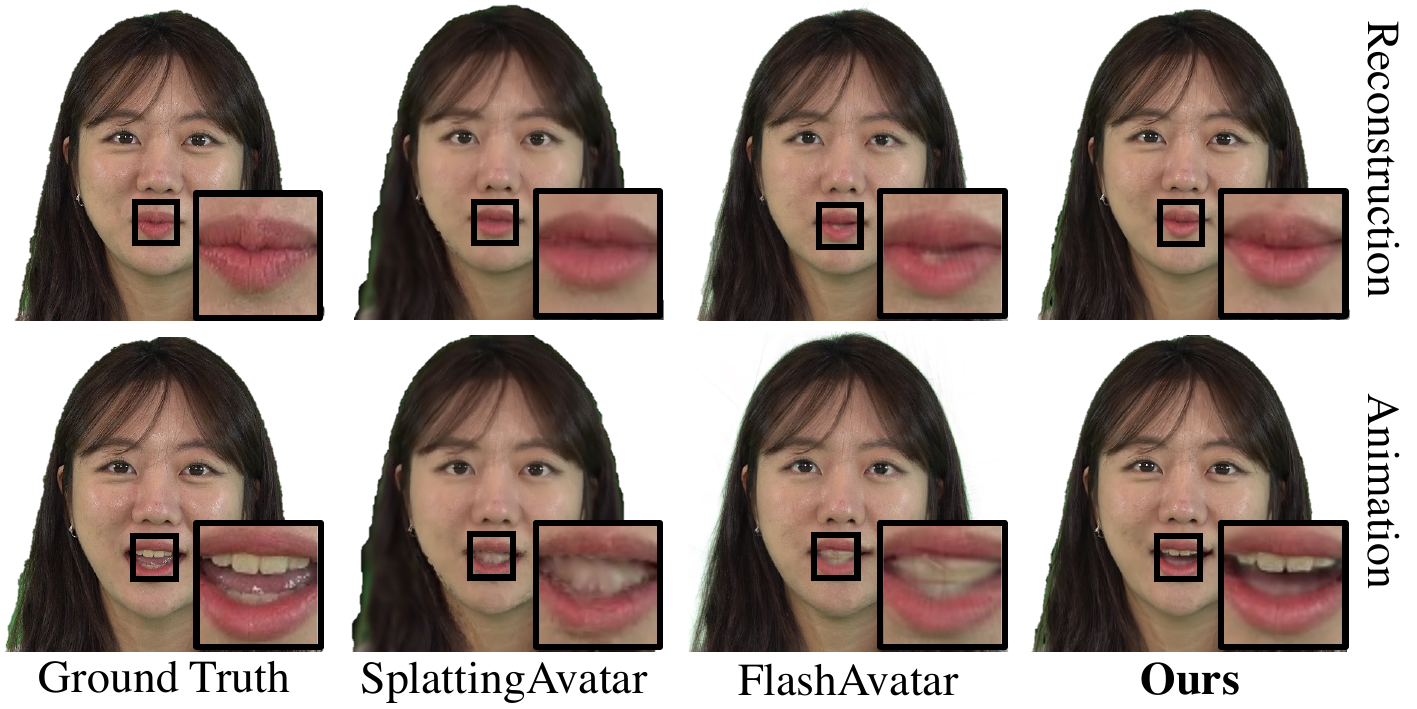}
\caption{Comparison of baseline performances.}
\label{fig:motivation_b}
\end{subfigure}
\begin{subfigure}{0.4\columnwidth}
\includegraphics[width=\columnwidth]{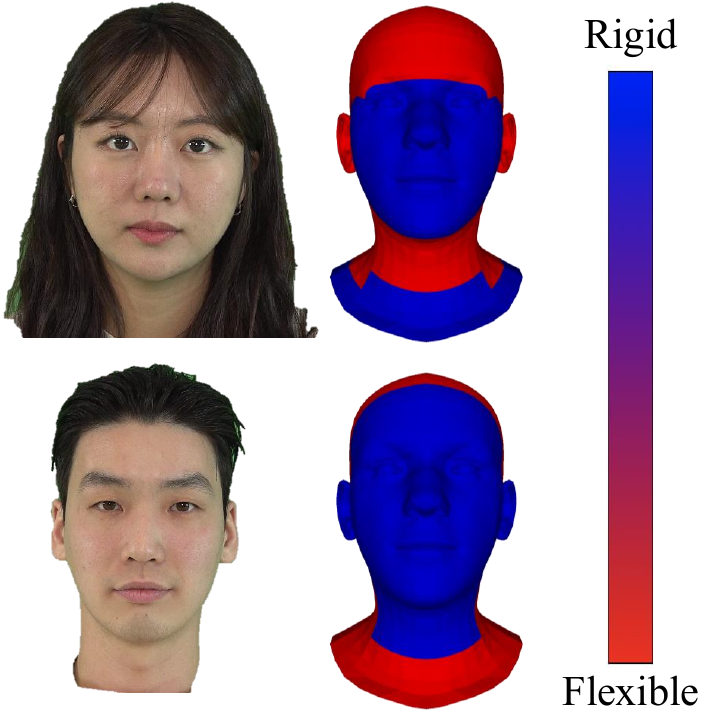}
\caption{Rigid and flexible sets.}
\label{fig:motivation_c}
\end{subfigure}
\end{center}
\vspace{-0.5cm}
\caption{
\textbf{Different distributions of the local mean per each facial region.}
(a) We plot the mean value of local means of Gaussians in each facial region.
Indeed, regions where the FLAME mesh cannot reconstruct the ground truth geometry, \eg, scalp, neck, and ear regions with long hair, show high mean values.
(b) However, existing models, \eg, SplattingAvatar and FlashAvatar, do not consider the aforementioned distribution differences, which show acceptable performance on reconstruction but not on the novel animation.
(c) In contrast, GeoAvatar considers the distribution differences across facial regions, \eg, \textit{bangs - flexible} and \textit{crew cut - rigid}.
}
\label{fig:motivation}
\vspace{-0.5cm}
\end{figure*}

To achieve this goal, the existing works simply apply offsets, \ie, local mean, to allow deviations for Gaussians~\cite{qian2024gaussianavatars, shao2024splattingavatar, zheng2024headgap, xiang2023flashavatar, dhamo2025headgas} while regularizing them through the na\"ive position loss based on the fixed threshold~\cite{qian2024gaussianavatars, zheng2024headgap}.
However, we observe a key finding that local mean distributions of Gaussians vary significantly across facial regions~\cite{qian2024gaussianavatars, shao2024splattingavatar, xiang2023flashavatar, zheng2024headgap, dhamo2025headgas}.
For example, less accurate regions of 3DMM fitting, \eg, scalp or ears, exhibit higher local mean values, while well-reconstructed regions, \eg, face or lips, show minimal deviation as shown in Figure~\ref{fig:motivation}(a).

Based on the key observation above, we point out that existing models apply the uniform regularization strategy for every facial region, not considering the distribution differences across facial regions.
The uniform regularization strategy refers to 1) no regularization \cite{xiang2023flashavatar, shao2024splattingavatar, dhamo2025headgas} or 2) mitigated regularization \cite{qian2024gaussianavatars, zheng2024headgap}, regardless of facial regions.
In other words, the uniform regularization strategy fails to apply stronger rigging regulation even in well-fitted regions, \eg, face, resulting in weakened correspondence between Gaussians and the mesh.
Indeed, in Figure~\ref{fig:motivation}(b), baselines yield acceptable reconstruction performance, whereas they generate severe artifacts during novel animations.

Moreover, several challenges arise when representing the mouth, due to the absence of fine details of mouth in 3DMMs \cite{li2017learning, gerig2018morphable} and high dynamics of mouth.
While recent works utilize mesh modification, \eg, adding extra faces by na\"ively connecting lip vertices \cite{xiang2023flashavatar} or lacking the mouth structure except for frontal teeth \cite{qian2024gaussianavatars}, the modified meshes still fail to represent the mouth structure properly, resulting in incomplete representations.
Furthermore, they utilize deformation \cite{xiang2023flashavatar, wang2023gaussianhead} to each Gaussian independently without considerations of mouth anatomy and dynamics.
Here, we point out that the mouth structure maintains the structural consistency within the same \textit{part}, \eg, upper teeth-palate and lower teeth-floor, during the animation.
On top of that, the less expressive and dynamic motions in the existing monocular video dataset~\cite{shao2024splattingavatar} decline the robustness and evaluation of the dynamic animation scenarios.


\noindent\textbf{Contributions.} We propose a novel adaptive \textit{\textbf{Geo}}metrical Gaussian Splatting framework for 3D head \textit{\textbf{Avatar}}, named \textit{\textbf{GeoAvatar}}.
Our method first segments Gaussians into two distinct categories: \textit{rigid} and \textit{flexible} sets.
Given a FLAME~\cite{li2017learning} mesh, Gaussians with small local means are classified as the rigid set, corresponding to regions where FLAME geometry closely aligns with the ground truth geometry. 
Conversely, Gaussians with large local means form the flexible set, representing regions where FLAME struggles to reconstruct the geometry accurately, requiring higher flexibility.
We present \textit{\textbf{A}daptive \textbf{P}re-allocation \textbf{S}tage} (\textit{\textbf{APS}}), an \textit{unsupervised} method to partition these sets.

In addition, we propose \textit{a novel mouth structure along with a deformation} to handle the high dynamics of the mouth.
First, we highlight that the mouth structure is absent in 3DMM~ \cite{li2017learning, palafox2021npms, gerig2018morphable} or still inadequate~\cite{qian2024gaussianavatars, xiang2023flashavatar}.
To this end, we construct not only frontal teeth \cite{qian2024gaussianavatars}, but also molar teeth, palate and floor, through the 3DMM mesh modification.
Moreover, with regarding the structural consistency, we propose a part-wise deformation on mouth regions, which deforms Gaussians within the same part with a consistent offset.
To the best of our knowledge, our work is the first attempt to adapt part-wise deformation considering the biological human anatomical structure.

Lastly, we design a tailored optimization scheme for GeoAvatar, introducing \textit{a novel regularization loss} for precise rigging Gaussians to corresponding FLAME faces. 
For the higher expressive and dynamic facial motions, we release a new monocular human face video dataset, \textit{DynamicFace}, featuring highly expressive facial motions. 
Our extensive experiments on existing benchmark datasets and DynamicFace demonstrate that GeoAvatar outperforms state-of-the-art methods qualitatively and quantitatively, excelling in both reconstruction and animation.
To sum up, The contributions of this work are as follows:

\begin{itemize}
    \item We introduce Adaptive Pre-allocation Stage (APS), an unsupervised method to segment rigid and flexible sets.
    \item We present a novel mouth structure and part-wise deformation, which significantly improves mouth textures.
    \item We introduce a novel regularization loss for precise rigging between Gaussians and 3DMM faces.
    \item We release a monocular human face video dataset, DynamicFace, which has a variety of expressions and poses.
\end{itemize}
\section{Related Work}
\label{rw}

\noindent\textbf{3D Gaussian Splatting with a 3D Morphable Model.} \
Gaussian Splatting \cite{kerbl20233d} has been widely used to reconstruct a static scene from 2D image sequences and corresponding 3D geometries.
The original Gaussian Splatting \cite{kerbl20233d} utilized Structure-from-Motion (SfM) \cite{ullman1979interpretation}, \eg, COLMAP \cite{schonberger2016structure}, to generate 3D point clouds from images \cite{fu2024colmapfree, lee2024deblurring}.
However, COLMAP has limitations for capturing 3D point clouds of dynamic objects \cite{saputra2018visual, bescos2018dynaslam}, \eg, human face \cite{park2021nerfies}.

A 3D Morphable Model (3DMM), \eg, FLAME \cite{li2017learning}, NPMs \cite{palafox2021npms}, or the Basel Face Model \cite{gerig2018morphable} has been widely used as a 3D prior for Gaussian Splatting in head avatar generation. 
For instance, one of the most widely used 3DMM models, FLAME \cite{li2017learning}, targets modeling the head and neck, allowing for the accurate capture of facial expressions and poses, including subtle movements of the jaw and eyes.

However, 3DMM inherently struggles to fully capture ground truth geometry due to several key limitations: 1) inherent low resolution of 3DMM, \eg, only 5023 vertices~\cite{li2017learning}, 2) fitting errors~\cite{kittler20163d, morales2021survey, kemelmacher20103d, dib2021towards}, and 3) lack of mouth structure, \eg, teeth~\cite{li2017learning, palafox2021npms, gerig2018morphable}.
These shortcomings prevent 3DMM from independently achieving accurate reconstruction of ground truth geometry.

\noindent\textbf{3D Head Avatar Reconstruction and Animation.} \
Recent methods~\cite{qian2024gaussianavatars, shao2024splattingavatar, xiang2023flashavatar, zheng2024headgap, dhamo2025headgas} introduce offsets for Gaussians, allowing them to deviate from the 3DMM mesh by a certain distance. 
This flexibility ensures that fine details, \eg, hair strands, the mouth structure, and regions prone to fitting errors, \eg, ears, can maintain their original identity and geometry, even with na\"ive FLAME geometry.

For instance, in SplattingAvatar~\cite{shao2024splattingavatar}, the model allows some degree of deviation between the generated Gaussians and the FLAME geometry to enhance flexibility during training.
In FlashAvatar~\cite{xiang2023flashavatar}, the model provides the offset for each Gaussian, which deforms them from the original FLAME mesh to the desirable location.

However, relying solely on a rendering loss during training often causes offsets to grow excessively large~\cite{qian2024gaussianavatars}. 
While this can improve the reconstruction quality, it compromises the correspondence between Gaussians and the underlying 3DMM structure with overly high positional freedom, leading to noticeable artifacts during animation~\cite{qian2024gaussianavatars}. 
To address this, existing models introduce a position loss that permits deviations through the local mean but regularizes based on the fixed threshold~\cite{qian2024gaussianavatars, zheng2024headgap}.
\section{Method}
\begin{figure*}[t]
\begin{center}
\includegraphics[width=1\textwidth]{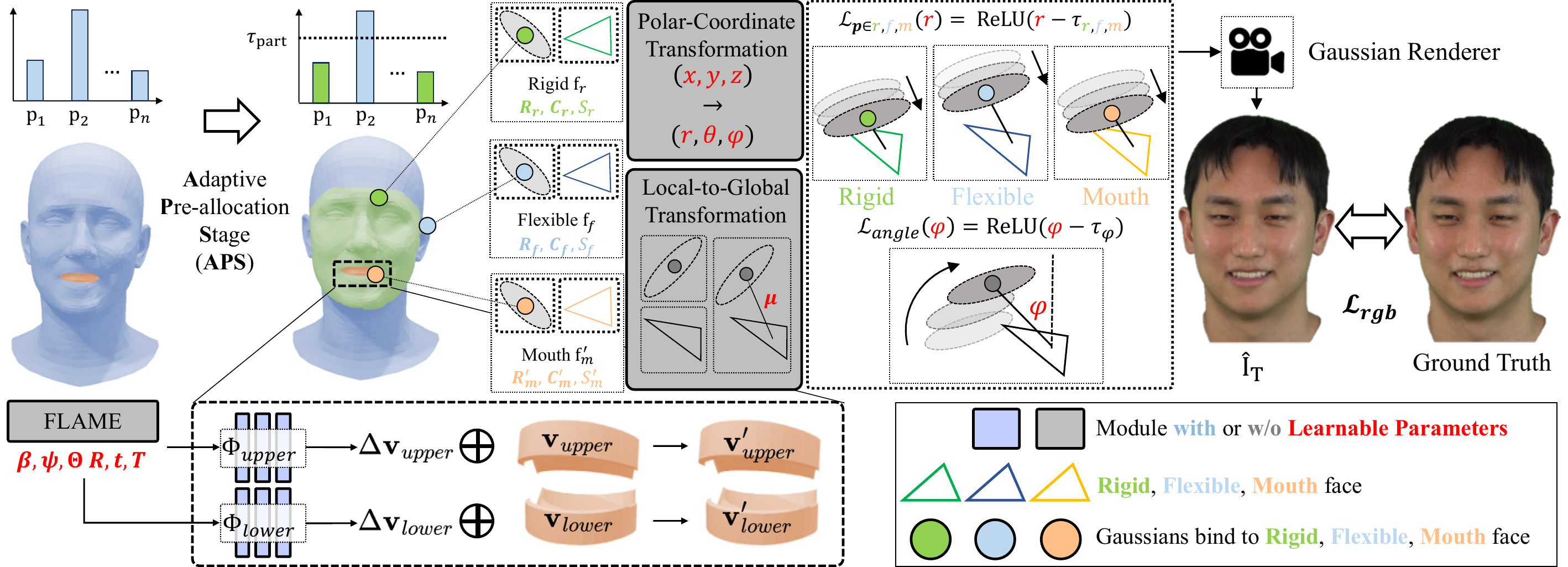}
\vspace{-0.7cm}
\end{center}
\caption{
\textbf{Structure of GeoAvatar.}
Our method segments 3DMM faces into three sets: rigid set, flexible set, and mouth structure set.
Adaptive Pre-allocation Stage segments rigid and flexible sets in an unsupervised way, by utilizing part-wise mean distance.
For the mouth structure set, we apply the deformation to its vertices part-wisely, \eg, upper and lower parts.
After deformation, we place local Gaussians into global coordinates through the local-to-global transformation, to yield the final rendering output through Gaussian renderer.
}
\label{fig:structure}
\vspace{-0.5cm}
\end{figure*}

Given the sequence of frames from the monocular video, we denote the frame corresponding to the timestep $T$ as $\text{I}_{T}$.
By utilizing FLAME tracker \cite{zielonka2022towards}, we extract FLAME parameters $\mathcal{F}=\{\boldsymbol\beta, \boldsymbol\psi, \boldsymbol\Theta, \textbf{R}, \textbf{t}\}$, consists of shape $\boldsymbol\beta\in\mathbb{R}^{300}$, expression $\boldsymbol\psi\in\mathbb{R}^{100}$, pose $\boldsymbol\Theta\in\mathbb{R}^{12}$, rotation $\textbf{R}\in\mathbb{R}^{3}$, and translation $\textbf{t}\in\mathbb{R}^{3}$, from $I_{T}$.
With $\mathcal{F}$, FLAME \cite{li2017learning} yields a mesh constructed by vertices $\textbf{V}\in\mathbb{R}^{5023\times3}$ and faces $\textbf{F}\in\mathbb{R}^{9976\times3}$.
Then, we utilize the given FLAME mask \cite{li2017learning} to segment FLAME faces into $n$ parts, where each set stands for a specific part of the face, \eg, face, ears, or lips.
In specific, we assign every face $\text{f}_{i}\in\textbf{F}$ to one of $n$ facial parts $\text{p}_{1}, \cdots \text{p}_{n}$.
During training, each face $\text{f}_{i}$ binds the set of Gaussians with index $j$, \ie, $G_{i}=\{\mathcal{G}_{i,j}|\mathcal{G}_{i,j} \text{ binds to } \text{f}_{i}\}$, where each Gaussian $\mathcal{G}$ can be expressed as below:
\begin{align}
    \mathcal{G} = \{\boldsymbol\mu, \textbf{r}, \textbf{s}, \textbf{c}, \alpha\},
\end{align}
where $\boldsymbol\mu\in\mathbb{R}^{3}$, $\textbf{r}\in\mathbb{R}^{4}$, $\textbf{s}\in\mathbb{R}^{3}$, $\textbf{c}\in\mathbb{R}^{3}$, and $\alpha$ stand for local mean, rotation, scale, color, and opacity, respectively.
We transform $\boldsymbol\mu=(x,y,z)$ into a polar coordinate, \ie, $\boldsymbol\mu=(r,\theta, \varphi)$, to apply the regularization proposed in Section \ref{method:3}.
Here, $r$ denotes the distance from the origin, $\theta$ and $\varphi$ denote the degree between $(x,y,z)$ and the positive $x$-axis, $z$-axis, respectively.

We segment faces in 3DMM into three sets: a rigid set $\textbf{F}_{r}$, a flexible set $\textbf{F}_{f}$, and a mouth structure set $\textbf{F}_{m}$.
We segment $\textbf{F}_{r}$ and $\textbf{F}_{f}$ in an unsupervised way, named \textit{Adaptive Pre-allocation Stage} (\textit{APS}) in Section \ref{method:1}, and set $\textbf{F}_{m}$ to be newly added faces for our modified FLAME, mentioned in Section \ref{method:2}.
To apply deformation on vertices that construct mouth structure faces, \ie, $\text{f}_{m}\in\textbf{F}_{m}$, we introduce a mouth deformation network $\Phi$ in Section \ref{method:2}, to yield the vertices offset $\Delta\textbf{v}_{m}$.
With deformed face $\text{f}_{m}'$ in $\textbf{F}_{m}$ and faces in $\textbf{F}_{r}$ or $\textbf{F}_{f}$, \ie, $\text{f}_{r}$, $\text{f}_{f}$, respectively, we obtain a new facial features, \eg, rotation $\textbf{R}_{\text{face}}$, center $\textbf{C}_{\text{face}}$, and scale $\textbf{S}_{\text{face}}$.
Utilizing facial features, $\mathcal{G}$ is transformed to the global space by following transformation $\textbf{T}$ \cite{qian2024gaussianavatars}:
\begin{align}
    \textbf{T}: \boldsymbol\mu, \textbf{s}, \textbf{r} \mapsto \textbf{S}_{\text{face}}\textbf{R}_{\text{face}}\boldsymbol\mu+\textbf{C}_{\text{face}}, \textbf{S}_{\text{face}}\textbf{s}, \textbf{R}_{\text{face}}\textbf{r}.
\end{align}
Finally, we go through the 3D Gaussian rendering process \cite{kerbl20233d} to yield the rendered output $\hat{\text{I}}_{T}$ as shown in Figure~\ref{fig:structure}.

\subsection{Adaptive Pre-allocation Stage}
\label{method:1}

We propose that faces of 3DMM can be distinguished into two sets besides pre-defined $\textbf{F}_{m}$: 1) faces that precisely reproduce the original 3D geometry and animation, \ie, rigid set $\textbf{F}_{r}$, 2) faces that fail to reproduce the 3D geometry but animates similarly with the ground truth, \ie, flexible set $\textbf{F}_{f}$.
Though FLAME offers masks that distinguish FLAME vertices part-wisely, \eg, eyes, nose, and lips, we need a criterion to classify each part $\text{p}_{k\in\{1,\cdots,n\}}$ to $\textbf{F}_{r}$ or $\textbf{F}_{f}$.
Since we do not have 3D priors at the beginning of the training, we set every face to $\textbf{F}_{r}$ except the mouth structure and optimize the model as described in Section \ref{method:3}, for $N$ steps.
After $N$ steps, we average the local mean of Gaussians that binds to $\text{f}_{i}$ and average it again part-wisely as below:
\begin{align}
    \text{Distance}(k) = \frac{1}{|\text{p}_{k}|}\sum_{\text{f}_{i}\in\text{p}_{k}}\frac{1}{|G_{i}|}\sum_{j\in{G_{i}}}{||\boldsymbol\mu_{i,j}||}.
\end{align}
Consequently, $\text{Distance}(k)$ can represent the mean distance between Gaussians belonging to $\text{p}_{k}$ and the mesh.
The large $\text{Distance}(k)$ means the model needs large flexibility to reproduce the original geometry on $\text{p}_{k}$, \ie, flexible set, and vice versa, \ie, rigid set.
We utilize the mean of $\text{Distance}(k)$ as a threshold $\tau_{part}$, and assign $\text{p}_{k}$ to $\textbf{F}_{r}$ if $\text{Distance}(k) < \tau_{part}$, and assign to $\textbf{F}_{f}$ if $\text{Distance}(k) > \tau_{part}$.

\subsection{Mouth Structure and Deformation}
\label{method:2}

\begin{figure}[h]
\begin{center}
\includegraphics[width=1\columnwidth]{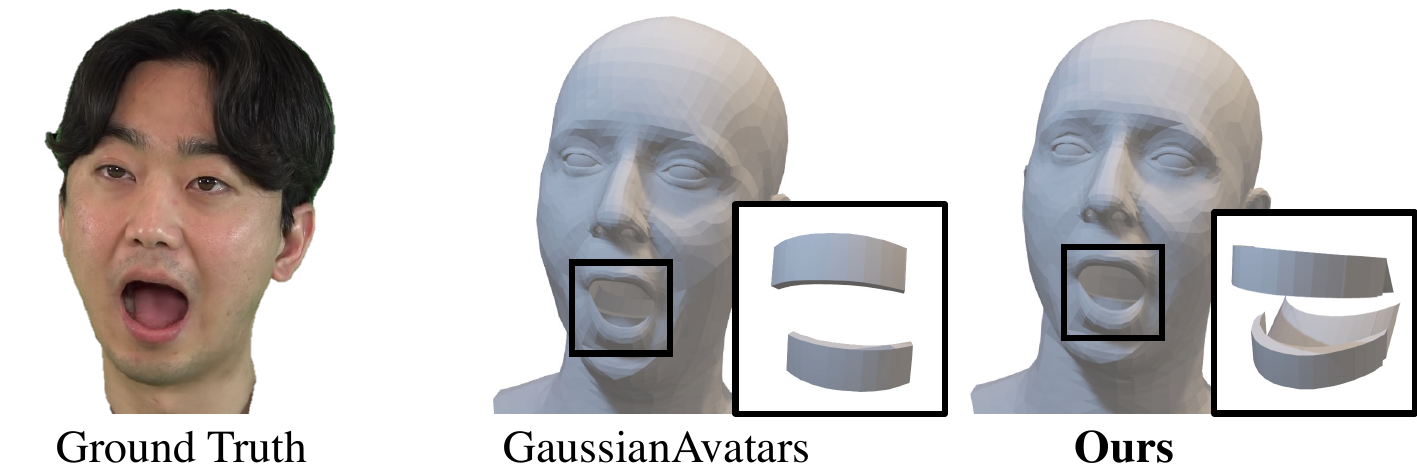}
\end{center}
\vspace{-0.5cm}
\caption{
\textbf{Visualization of mouth structure.}
We compare our mouth structure with GaussianAvatars \cite{qian2024gaussianavatars} with the widely open-mouth frame.
The structure of GaussianAvatars only covers the frontal teeth and not enough to cover the molar teeth.
On the contrary, ours has not only the frontal and molar teeth but also palate and floor structures for more elaborate representation.
}
\label{fig:flame}
\vspace{-0.3cm}
\end{figure} 

\noindent Arguably, since the mouth part is the most dynamic part among all facial regions, improving the generation quality of the mouth is crucial to enhance the animation performance.
However, FLAME lacks the geometry of the mouth structure, which degrades the generation performance on the existing models \cite{qian2024gaussianavatars, cho2024gaussiantalker, xiang2023flashavatar}.
To this end, we add the mouth structure $\textbf{F}_{m}$ inside FLAME.
Motivated by GaussianAvatars \cite{qian2024gaussianavatars}, we utilize the lip ring meshes to construct teeth.
However, we empirically find that this method can only cover the frontal teeth, while the molar teeth geometry is still empty.
To offer more precise geometrical priors, we extend the teeth structure to cover the molar teeth and additionally add the palate and floor of the mouth structure.
In Figure \ref{fig:flame}, we compare our mouth structure with GaussianAvatars \cite{qian2024gaussianavatars}, which shows more elaborate and realistic.
We elucidate the detailed process in Appendix \ref{app:structure}.

Though we incorporate the mouth structure within the FLAME, 3DMM tracking cannot accurately capture the subtle mouth animations.
Therefore, to mitigate this error, models utilize the deformation \cite{xiang2023flashavatar, wang2023gaussianhead} per Gaussian.
However, even during the dynamic movement of the mouth, the relative locations within the upper part, \eg, upper teeth and palate, and the lower part, \eg, lower teeth and floor, remain consistent.
In light of this observation, we suggest \textit{part-wise deformation}, which maintains the relative locations within the same part.
In specific, we divide the mouth structure into two parts, \eg, upper part $\textbf{v}_{upper}$ for upper teeth and palate vertices, and the lower part $\textbf{v}_{lower}$ for lower teeth and floor vertices.
With offsets for each part, \ie, $\Delta \textbf{v}_{upper}$ and $\Delta \textbf{v}_{lower}$, we deform vertices in each part simultaneously:

\begin{align}
    \textbf{v}\prime_{upper} = \textbf{v}_{upper} + \Delta \textbf{v}_{upper}, \\
    \textbf{v}\prime_{lower} = \textbf{v}_{lower} + \Delta \textbf{v}_{lower}.
\end{align}

Consequently, the Gaussians rigged to each mouth part are deformed concurrently along with mouth structures.
We obtain $\Delta \textbf{v}_{upper}$ and $\Delta \textbf{v}_{lower}$ via separate deformation networks with the same structure, \ie, $\Phi_{upper}$ and $\Phi_{lower}$, respectively.
In each $\Phi$, we utilize FLAME parameters related to mesh animation \cite{cudeiro2019capture}, \eg, $\boldsymbol\psi$ and $\boldsymbol\Theta$, along with timestep $T$, as inputs.
While recent deformation models \cite{wang2023gaussianhead, xiang2023flashavatar, xu2024gaussian} only utilize FLAME parameters or landmarks, we empirically find that it is insufficient to address the subtle misalignment per each frame.
Therefore, we utilize $T$ with FLAME parameters for timestep-wise mesh correction:
\begin{align}
    \Delta \textbf{v} = \Phi(\boldsymbol\psi, \boldsymbol\Theta, \gamma(T)),
\end{align}
where $\gamma$ denotes the positional encoding \cite{mildenhall2021nerf, vaswani2017attention}.
In inference, since we do not need the FLAME correction, we simply set $T=0$.

\subsection{Optimization Scheme}
\label{method:3}
\vspace{-0.3cm}
\begin{figure}[h]
\begin{center}
\includegraphics[width=1\columnwidth]{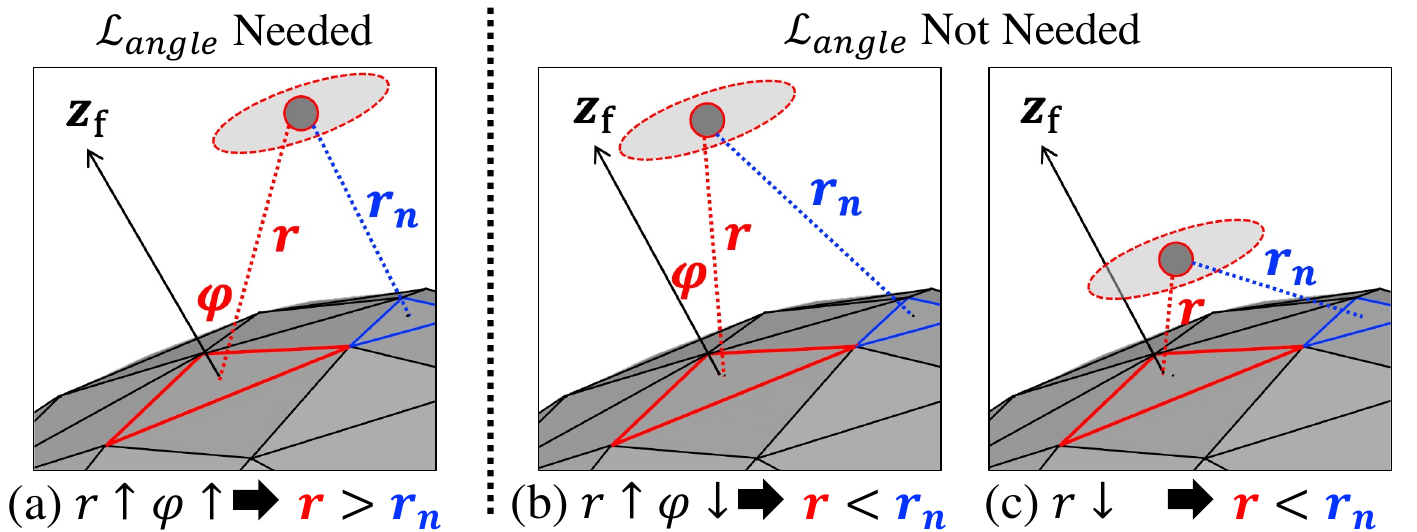}
\end{center}
\vspace{-0.5cm}
\caption{
\textbf{Strategy for rigging Gaussians correctly.}
(a) With high values of $r$ and $\varphi$, Gaussians may be positioned far from the rigged triangle, \ie, denoted as red, increasing the likelihood of representing a different part from the one assigned to the red triangle.
(b) While maintaining the high $r$, \eg, flexible set, we can keep Gaussians not to be closer to the neighborhood triangle by regularizing $\varphi$.
(c) In the case when $r$ is small, \eg, rigid set, we do not need to regularize $\varphi$.
}
\label{fig:spherical}
\vspace{-0.3cm}
\end{figure} 

\noindent In addition to the loss function that compares the generated and original images, models that utilize 3DMM often introduce regularization terms \cite{qian2024gaussianavatars} or training strategies \cite{shao2024splattingavatar} to preserve the geometric consistency between the 3DMM mesh and Gaussians.
This approach is essential for the following two reasons:
First, to ensure that the Gaussian-based avatar exhibits similar animations as the 3DMM, models should maintain a rough alignment between the geometry of the Gaussians and the 3DMM mesh.
To achieve this, we impose a regularization on the local mean of the Gaussians, regularizing the center of each Gaussian to be located in proximity to the center of its corresponding mesh face.
We use the radius of $\boldsymbol\mu$, \ie, $r$, obtained by polar coordinate transformation:
\begin{align}
\label{eq:reg_surf}
   \mathcal{L}_{p}(r) = \text{ReLU}(r - \tau_{p}),
\end{align}
where $p$ stands for the set, \ie, rigid $r$, flexible $f$, or mouth $m$.
Specifically, we set $\tau_{r}=0.1$, $\tau_{f}=2.0$ to adjust the regularization strength flexibly.
Lastly, in the case $\tau_{m}$, we can assume $\textbf{F}_{m}$ reflects the ground truth geometry properly through the deformation.
Therefore, we apply strict regularization loss term for $\textbf{F}_{m}$ same as $\textbf{F}_{r}$, \ie, $\tau_{m}=\tau_{r}$.

Second, the other purpose of regularization is to ensure the proper assignment of Gaussian animations.
For instance, a Gaussian representing the lip skin but rigged to a mesh face representing the nose can cause artifacts in the animation.
This misalignment occurs because the position of the Gaussian does not closely align with the intended facial region, causing undesired deformation in nearby areas.
Consequently, it is desirable to keep the distance between $\mathcal{G}_{i,j}$ and $\text{f}_{i}$ be smaller than the distance between $\mathcal{G}_{i,j}$ and the neighborhood triangle, \ie, $r_n$.
In the case when $r$ is small, we can keep $r < r_{n}$ even with large $\varphi$, as shown in Figure \ref{fig:spherical}(c).
However, to facilitate flexibility in geometry, we use a larger $\tau_{f}$, which allows Gaussians to have large $r$.
As illustrated in Figure \ref{fig:spherical}(a), this may lead to $r > r_{n}$ when $\varphi$ is large.
To this end, we introduce an additional loss term:
\begin{align}
\label{eq:reg_sph}
    \mathcal{L}_{angle}(\varphi) = \mathbbm{1}_{{r} > \tau_{r}} \text{ReLU}(\varphi\ - \tau_{\varphi}),
\end{align}
which regularizes $\varphi$ if $r>\tau_{r}$.
We set $\tau_{\varphi} = 45^\circ \approx 0.78\text{ rad}$.
Finally, our regularization loss $\mathcal{L}_{reg}$ is following:
\begin{align}
   \mathcal{L}_{reg}(\boldsymbol{\mu}) = \sum_{p\in\{r, f, m\}}\sum_{\text{f}_{i}\in \textbf{F}_p}\sum_{\mathcal{G}_{i,j}\in G_{i}}\mathcal{L}_{p}(r_{i,j}) + \mathcal{L}_{angle}(\varphi_{i,j}).
\end{align}

Combining proposed loss terms with the rendering loss $\mathcal{L}_{rgb} = (1-\lambda)\mathcal{L}_{1}+\lambda\mathcal{L}_{\text{D-SSIM}}$ \cite{kerbl20233d}, our final optimization loss $\mathcal{L}$ is defined as below:

\begin{align}
   \mathcal{L} = \mathcal{L}_{rgb}(I,\hat{I}) + \mathcal{L}_{reg}(\boldsymbol{\mu}).
\end{align}



\section{Experiments}

\begin{figure*}[t]
\begin{center}
\includegraphics[width=1\textwidth]{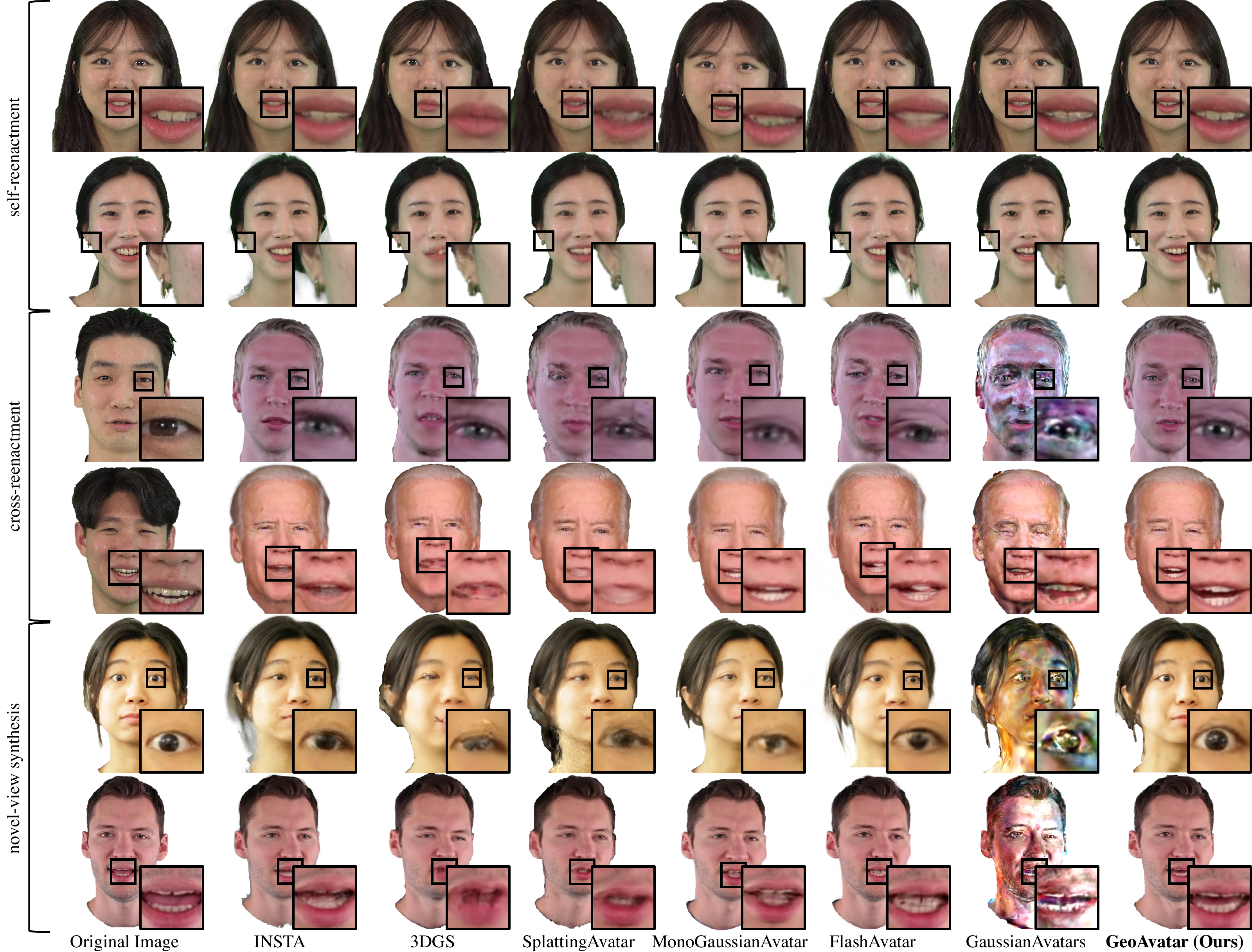}
\end{center}
\vspace{-0.5cm}
\caption{
\textbf{Self- and cross-reenactment and novel-view synthesis results.}
We compared our method with baselines on various scenarios, \ie, self- and cross-reenactment and novel-view synthesis.
In self-reenactment, ours showed robust generation results while maintaining high texture quality.
In cross-reenactment, ours presented highly expressive and robust results similar to source actors, while the others showed artifacts and blurred details with low similarity of driving expressions.
In novel-view synthesis, ours successfully reproduced the animation without artifacts.
On the other hand, baselines struggled to reproduce desirable animations or to generate without artifacts.
}
\label{fig:monocular}
\vspace{-0.5cm}
\end{figure*}

In this section, we briefly introduce datasets and baselines.
Then, we compared the performance on self- and cross-reenactment, and novel-view synthesis, while showing the effectiveness of each method through ablation studies.

\subsection{Setup}
For thorough evaluations, we utilized various datasets including our newly proposed dataset, \textbf{DynamicFace}. Following are the brief descriptions of each dataset:

\noindent\textbf{DynamicFace.}
To focus on generating dynamic facial movements, we collected a dataset, consisting of 2-3 minutes of monocular RGB videos featuring 10 actors.
Each actor was instructed to perform a range of facial expressions for the construction of a dataset that can facilitate the development of robust models of facial movements.
Additionally, to ensure coverage of various viewpoints with a single camera, actors were instructed to slowly nod their heads while holding different facial expressions.
We elucidated details for collecting DynamicFace in Appendix \ref{app:dataset}.

\noindent\textbf{SplattingAvatar.}
We utilized datasets provided by SplattingAvatar \cite{shao2024splattingavatar}, which consists of 10 monocular videos from INSTA \cite{zielonka2023instant}, NHA \cite{grassal2022neural}, IMAvatar \cite{zheng2022avatar}, and NerFace \cite{gafni2021dynamic}.

\noindent\textbf{NeRSemble.}
Since NeRSemble \cite{kirschstein2023nersemble} provides a 16-view dataset of various human facial animations, we utilized it to compare our model with the multi-view avatar generation model in Section \ref{res:multi}.
We utilized videos of 12 actors from the NeRSemble dataset for a thorough comparison.

\noindent\textbf{Baselines.}
We compared our method with the state-of-the-art NeRF-based model, \ie, INSTA \cite{zielonka2023instant} and various outperforming Gaussian Splatting-based models, \ie, SplattingAvatar \cite{shao2024splattingavatar}, MonoGaussianAvatar \cite{chen2024monogaussianavatar}, FlashAvatar \cite{xiang2023flashavatar}, GaussianAvatars \cite{qian2024gaussianavatars}, and including 3DGS \cite{kerbl20233d} itself.
Though ours targets the monocular video setting, we included the multi-view video-based model, \ie, GaussianAvatars \cite{qian2024gaussianavatars}, due to its notable performance.
Baseline details are explained in Appendix \ref{app:baseline}.

\subsection{Comparison on Monocular Video Dataset}

\begin{table*}[ht]
\centering\footnotesize
\resizebox{\textwidth}{!}{%
\begin{tabular}{l|cccccccc}
\toprule
Model & INSTA \cite{zielonka2023instant} & 3DGS \cite{kerbl20233d}  & SplattingAvatar \cite{shao2024splattingavatar} & MonoGaussainAvatar \cite{chen2024monogaussianavatar} & FlashAvatar \cite{xiang2023flashavatar} & GaussianAvatars \cite{qian2024gaussianavatars} & $\text{GaussianAvatars}_{0}$ \cite{qian2024gaussianavatars} &
\textbf{GeoAvatar (Ours)} \\
\midrule
Dataset & \multicolumn{7}{c}{SplattingAvatar \cite{shao2024splattingavatar}} \\
\midrule
MSE \scriptsize($10^{-3}$) $\downarrow$ & \phantom{0}2.429\stdv{2.02} &
\phantom{0}2.164\stdv{2.14} &
\phantom{0}3.132\stdv{0.83} &
\phantom{0}1.895\stdv{1.73} &
\phantom{0}2.239\stdv{2.63} &
\phantom{0}1.679\stdv{1.53} & \phantom{0}\underline{1.237\stdv{0.80}} & \textbf{\phantom{0}0.884\stdv{0.72}} \\
PSNR $\uparrow$ &
28.083\stdv{2.50} &
27.543\stdv{2.95} &
25.333\stdv{1.19} &
28.813\stdv{2.95} &
\underline{29.306\stdv{2.85}} & 
29.124\stdv{3.51} & 
29.686\stdv{2.91} & \textbf{32.635\stdv{2.88}} \\
SSIM $\uparrow$ &
\phantom{0}0.938\stdv{0.02} &
\phantom{0}0.923\stdv{0.03} &
\phantom{0}0.933\stdv{0.02} & 
\phantom{0}0.937\stdv{0.02} &
\phantom{0}\underline{0.943\stdv{0.02}} &
\phantom{0}0.938\stdv{0.03} & 
\phantom{0}0.934\stdv{0.02} &
\textbf{\phantom{0}0.965\stdv{0.02}} \\
LPIPS \scriptsize($10^{-1}$) $\downarrow$ & \phantom{0}0.678\stdv{0.23} &
\phantom{0}1.019\stdv{0.42} &
\phantom{0}0.588\stdv{0.12} & 
\phantom{0}0.733\stdv{0.38} &
\phantom{0}\underline{0.444\stdv{0.17}} & 
\phantom{0}0.494\stdv{0.28} & 
\phantom{0}0.529\stdv{0.16} & \textbf{\phantom{0}0.367\stdv{0.17}} \\
\midrule
Dataset & \multicolumn{7}{c}{DynamicFace (\textbf{Ours})} \\
\midrule
MSE \scriptsize($10^{-3}$) $\downarrow$ & \phantom{0}1.545\stdv{0.59} & \phantom{0}1.603\stdv{0.77} & \phantom{0}1.426\stdv{0.52}
& \phantom{0}1.618\stdv{0.51}
& \phantom{0}1.811\stdv{0.61} & 
\phantom{0}1.268\stdv{0.71} & \phantom{0}\underline{0.744\stdv{0.39}} & \textbf{\phantom{0}0.612\stdv{0.35}} \\
PSNR $\uparrow$ & 28.688\stdv{1.96} & 28.547\stdv{1.94}
& 28.843\stdv{1.63}
& 28.196\stdv{1.32} 
& 27.780\stdv{1.36}
& 29.641\stdv{2.24}
& \underline{31.026\stdv{2.09}} & \textbf{32.760\stdv{1.99}} \\
SSIM $\uparrow$ & \phantom{0}0.888\stdv{0.03} 
& \phantom{0}0.874\stdv{0.03}
& \phantom{0}0.869\stdv{0.03}
& \phantom{0}0.886\stdv{0.03}
& \phantom{0}0.874\stdv{0.03}
& \phantom{0}0.879\stdv{0.04}
& \phantom{0}\underline{0.904\stdv{0.02}} & \textbf{\phantom{0}0.919\stdv{0.02}} \\
LPIPS \scriptsize($10^{-1}$) $\downarrow$
& \phantom{0}1.289\stdv{0.28}
& \phantom{0}1.699\stdv{0.34}
& \phantom{0}1.378\stdv{0.21}
& \phantom{0}1.321\stdv{0.17}
& \phantom{0}\underline{0.745\stdv{0.10}}
& \phantom{0}0.801\stdv{0.20}
& \phantom{0}0.776\stdv{0.18} & \textbf{\phantom{0}0.660\stdv{0.14}} \\
\bottomrule
\end{tabular}
}
\caption{
\textbf{Quantitative comparison on SplattingAvatar \cite{shao2024splattingavatar} and DynamicFace datasets.}
We utilized 10 subjects from each dataset and denoted the mean value with standard deviation from each dataset.
\textbf{Bold} indicates the best and \underline{underline} indicates the second.
}\label{tab:main}
\vspace{-0.5cm}
\end{table*}

In Figure \ref{fig:monocular}, we compared our method with state-of-the-art baselines in self- and cross-reenactment and novel-view synthesis scenarios.
When presented with unseen animations near the mouth, \eg, the first row, the baselines exhibited severe artifacts.
On the contrary, with the robust mouth modification and deformation schemes, ours showed clear results without artifacts.
In the second row, ours showed the highest resolutions on the accessory part, due to the flexible relaxation of the geometry through APS.

Then, we showed the cross-reenactment and novel-view synthesis results of each model.
Since GaussianAvatars was originally a multi-view video-based model, it showed severe artifacts with a monocular-video dataset.
As seen in the third and fourth rows, ours showed superiority in the cross-reenactment scenario, while other baselines showed low-textured results with less accurate expressions.
Moreover, though baselines generated coarse appearances well, they suffered from generating the original motion, leading to artifacts and blurred details as shown in the fifth and sixth rows.
In contrast, ours showed robust and highly textured results.
Please refer to Appendix \ref{app:experiments} for additional results.

In Table \ref{tab:main}, we quantitatively compared the self-reenactment results of each model on both monocular video datasets, \eg, DynamicFace and SplattingAvatar.
For the fair comparison, we utilized every 10 subjects in each dataset and set the last 350 frames of each video as test sequences, as \cite{shao2024splattingavatar} and \cite{zielonka2023instant} did.
Our model outperformed the baselines on both datasets, with notable margins across various metrics, \eg, MSE, PSNR, SSIM, and LPIPS.
Note that, we design the test dataset in DynamicFace with challenging scenarios, \eg, dynamic talking scenarios, to evaluate the model performance more accurately. 
Indeed, the models show notably worse results on the human-aligned metric, \eg, LPIPS \cite{zhang2018unreasonable}, in DynamicFace compared to SplattingAvatar Dataset.
We show additional quantitative results on self and cross reenactment in Appendix \ref{app:quantitative}.

\subsection{Comparison on Multi-View Video Dataset}
\label{res:multi}
\begin{figure}[t]
\begin{center}
\includegraphics[width=1\columnwidth]{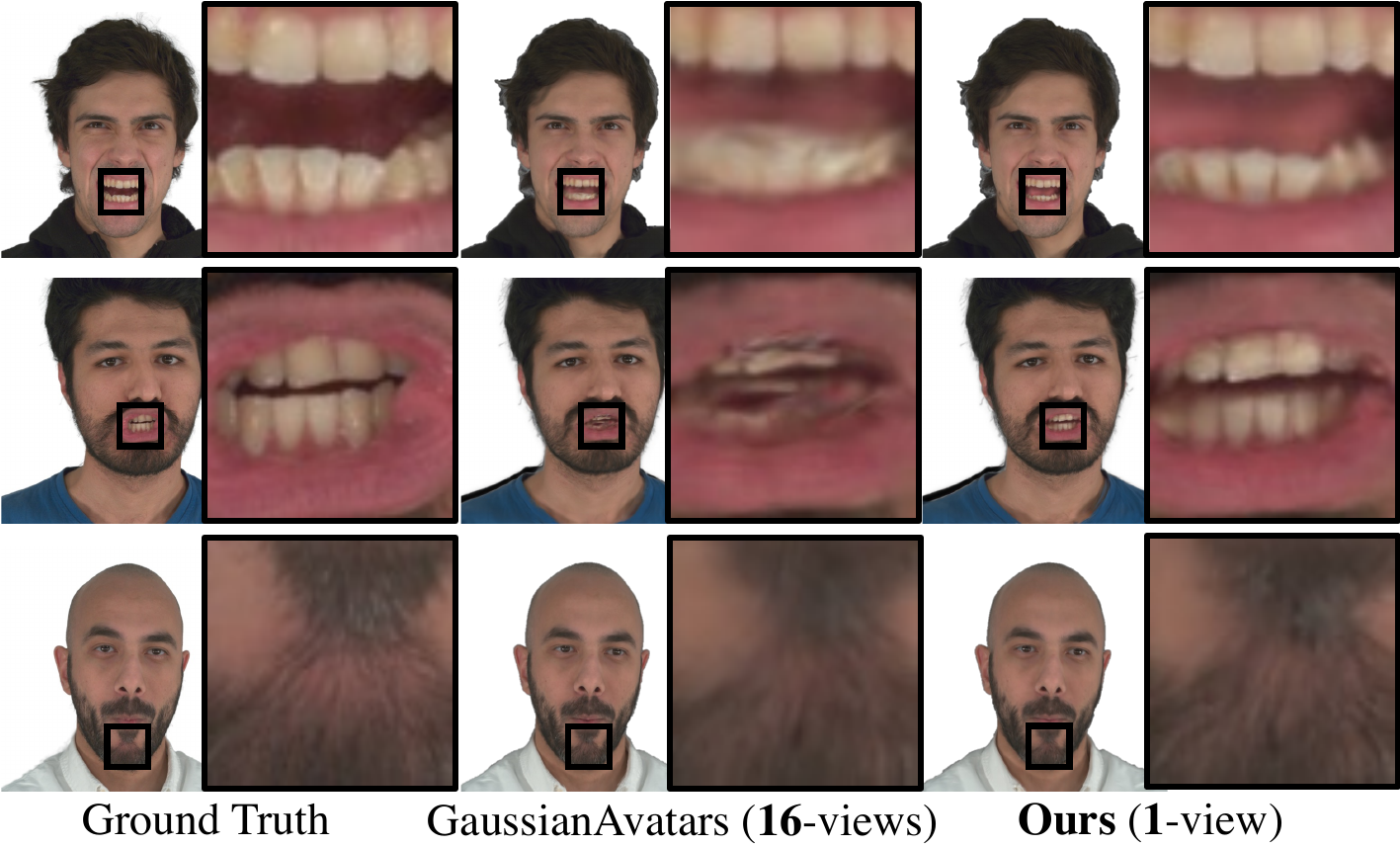}
\end{center}
\vspace{-0.5cm}
\caption{
\textbf{Qualitative comparison on the multi-view dataset.}
We compared the result between ours trained on a single view, and GaussianAvatars trained on 16 views.
Though we utilized only a single view, ours showed remarkably better reproduction on teeth, while improving the texture besides the mouth.
}
\vspace{-0.5cm}
\label{fig:multi_view}
\end{figure}
\begin{table}[ht]
\centering\footnotesize
\resizebox{\columnwidth}{!}{%
\begin{tabular}{l|cc|c}
\toprule
& \makecell{GaussianAvatars\\(16-views)} & \makecell{\textbf{Ours}\\(1-view)} & Gap (\%) \\
\midrule
MSE \scriptsize($10^{-3}$) $\downarrow$ & \textbf{\phantom{0}2.483\stdv{2.16}} & \phantom{0}2.514\stdv{2.18} & -1.247 \\
PSNR $\uparrow$ & \textbf{27.829\stdv{1.59}} & 27.782\stdv{1.60} & -0.169 \\
SSIM $\uparrow$ & \phantom{0}0.877\stdv{0.04} & \textbf{\phantom{0}0.882\stdv{0.04}} & +0.617 \\
LPIPS \scriptsize($10^{-1}$) $\downarrow$ & \phantom{0}1.073\stdv{0.34} & \textbf{\phantom{0}0.969\stdv{0.32}} & +9.711 \\
\bottomrule

\end{tabular}
}
\caption{
\textbf{Quantitative comparison on the multi-view dataset.}
We utilized 12 actors in the NeRSemble dataset and denoted the mean and standard deviation.
With only utilizing a single view, ours showed comparable results with GaussianAvatars.
In specific, ours showed a marginal difference on MSE, PSNR, and SSIM, while showing a notable gain on LPIPS, \ie, 9.7\%.
\textbf{Bold} indicates the best and \underline{underline} indicates the second.
}
\vspace{-0.5cm}
\label{tab:multi}
\end{table}

Though multi-view settings can improve the FLAME fitting results through batch-wise optimization, we conjecture that it is not enough to reflect detailed geometry and subtle movements inside the mouth.
To prove this, we conducted the experiment on the NeRSemble dataset, which utilizes multi-view FLAME fitting and calibrated camera parameters to provide accurate FLAME fitting results.
Using NeRSemble preprocessing, we compared ours with the state-of-the-art multi-view video-based avatar, \ie, GaussianAvatars.
Since our model is a monocular video-based model, we utilized only a single view to train ours, while GaussianAvatars employed every view, \ie, 16 views. 

First, we qualitatively compared ours and GaussianAvatars as in Figure \ref{fig:multi_view}.
Though the dataset provides elaborated FLAME fitting results, GaussianAvatars still suffered from artifacts inside the mouth.
On the other hand, our model effectively reduced artifacts inside the mouth by utilizing \textit{only single-view} images.
Moreover, even besides the mouth, ours showed sharper textures than GaussianAvatars, which proves that our flexible utilization of the geometry is useful even when the FLAME fitting result is relatively more accurate than the monocular video scenario.
In Table \ref{tab:multi}, we emphasized that ours showed comparable results on MSE and PSNR, and better results on SSIM and LPIPS than GaussianAvatars, by utilizing only a \textit{single} view for the training.
Specifically, ours outperformed on LPIPS metric with the notable margin, \ie, 9.7\%, which is closely related to human perception quality \cite{zhang2018unreasonable}.

\subsection{Ablation Study}
\begin{figure*}[t]
\begin{center}
\includegraphics[width=1\textwidth]{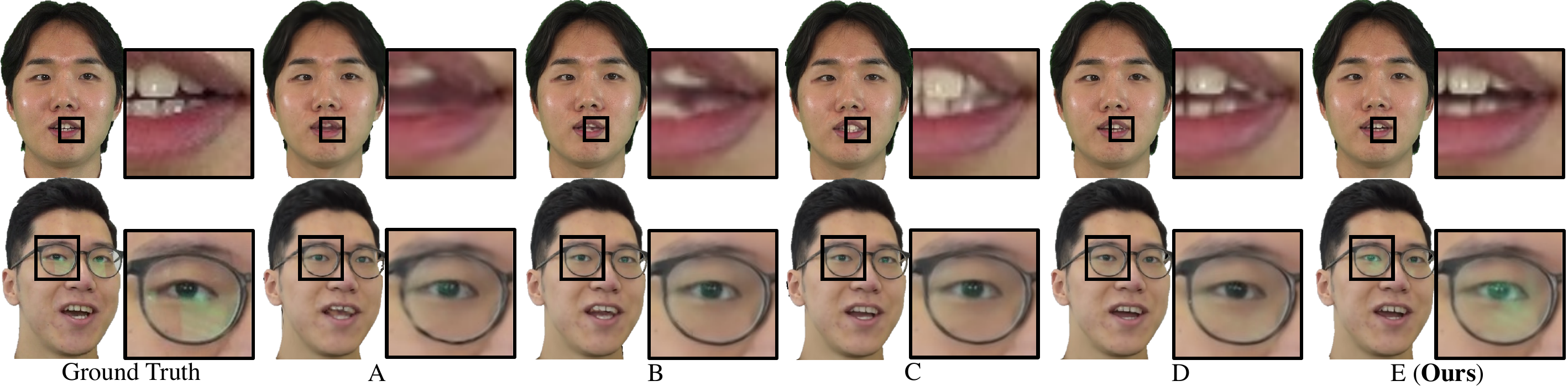}
\vspace{-1.0cm}
\end{center}
\caption{
\textbf{Qualitative ablation results.}
We compared each ablation configuration qualitatively.
With applying APS, \ie, B, the resolution of output improved notably.
With FLAME teeth and mouth deformation, \ie, C and D, the teeth generation quality improved remarkably.
Finally, By adding $\mathcal{L}_{angle}$, \ie, E, we can wipe out existing artifacts and obtained more robust results.
Best viewed zoom-in.
}
\vspace{-0.5cm}
\label{fig:ablation}
\end{figure*}

\noindent To demonstrate the effectiveness of each method proposed in GeoAvatar, we set the configurations as follows:
First, A is same with the baseline \cite{qian2024gaussianavatars} except for setting spherical harmonics degree of Gaussians from three to zero, $\text{GaussianAvatars}_{0}$, to fit better for the monocular video-based scenario.
Then B was trained by adding APS to A, while C utilized FLAME mouth modification from B.
For D, we added the mouth part-wise deformation.
Finally, by adding $\mathcal{L}_{angle}$ to D, we constructed E, \ie, GeoAvatar.

\begin{table}[t]
\centering\footnotesize
\resizebox{\columnwidth}{!}{%
\centering
\begin{tabular}{ll|cccc}
\toprule
& Configuration & \makecell{MSE \\ \scriptsize($10^{-3}$) $\downarrow$} & PSNR $\uparrow$ & SSIM $\uparrow$ & \makecell{LPIPS \\ \scriptsize($10^{-1}$) $\downarrow$} \\
\midrule
A & Baseline & 0.991 & 30.356 & 0.919 & 0.653 \\
\midrule
B & + APS & 0.905 & 30.717 & 0.930 & 0.572 \\
C & + FLAME mouth & 0.802 & 32.273 & 0.941 & 0.548 \\
D & + Part-wise deformation & \textbf{0.733} & \textbf{32.751} & \underline{0.941} & \underline{0.519} \\
E & + $\mathcal{L}_{angle}$ (\textbf{Ours}) & \underline{0.748} & \underline{32.697} & \textbf{0.942} & \textbf{0.513} \\
\bottomrule
\end{tabular}
}
\caption{
\textbf{Quantitative ablation results.}
We compared the performance of each proposed method from the baseline, which only utilized the method for rigging 3D Gaussians along with 3DMM \cite{qian2024gaussianavatars}.
Our proposed method indeed improved results quantitatively, especially for the human-aligned metric, \ie, LPIPS.
}\label{tab:ablation}
\vspace{-0.5cm}
\end{table}

First, B showed a notably lower LPIPS than A, while showing comparable results on other metrics.
We attributed this to the effective regularization by setting the rigid set, which improved the human-aligned metric, \ie, LPIPS.
Moreover, by separating the region that needs large flexibility, \ie, flexible set, from the rigid set through APS, B obtained better performance even on metrics that are related to the reconstruction, \eg, MSE and PSNR.
Indeed, in Figure \ref{fig:ablation}, B generated sharper teeth and eyeglasses than A.

Adding the mouth structure to the mesh, \ie, C, further enhanced the quality of the teeth, but it still generated a blurred result and a wrong teeth geometry.
The mouth structure showed a synergistic effect when applied with $\Phi$, which effectively led teeth to resemble the ground truth.

Finally, adding $\mathcal{L}_{angle}$ in E, rigging of Gaussians became more robust, yielding more accurate results.
For instance, in Figure \ref{fig:ablation}, E generated teeth without any artifacts inside the mouth, \ie, the first row, while robustly representing the light reflection on the eyeglasses, \ie, the second row.
We show additional ablations on initializations, APS, and deformation field more elaborately by isolating each module in Appendix \ref{app:abl}.
Moreover, we visualize the distribution of rigid and flexible sets in Appendix \ref{app:vis}.
\section{Discussion}
\noindent\textbf{Limitations and Future Works.} Although GeoAvatar showed superior and robust 3D head avatar generation compared to the existing baselines, it still has following issues: 1) Since our model lacks hair and cloth modeling, combining recent hair and clothing methods~\cite{luo2024gaussianhair, rong2024gaussian, lin2024layga} is a promising research.
2) Currently, since our method is not feasible for relighting avatars~\cite{he2024diffrelight}, considering lighting conditions has more room to be interesting in further steps.

\noindent\textbf{Negative Social Impacts.}
Our advanced 3D head avatar framework could be misused to create highly realistic fake identities or deepfake content, along with the misuse of the released dataset, DynamicFace. 
To mitigate these risks, we emphasize the importance of users following ethical guidelines, ensuring data security, and reflecting on the broader ethical implications in applying our technology.
\section{Conclusion}
In this work, we proposed GeoAvatar, a framework for adaptive geometrical Gaussian Splatting in 3D head avatar generation. 
We introduced Adaptive Pre-allocation Stage (APS), along with a novel mouth structure and part-wise deformation, and new regularization loss terms for high-quality reconstruction and animation.
Additionally, we release DynamicFace, a monocular video dataset featuring highly expressive facial motions.
Extensive experiments showed GeoAvatar surpasses state-of-the-art methods in both reconstruction and animation quality.
We look forward to the active utilization of GeoAvatar in future works and applications within related domains.
\section{Acknowledgements}
This work was supported by the Korea Creative Content Agency (KOCCA) grant funded by the Ministry of Culture, Sports and Tourism of the Republic of Korea (RS-2024-00441174) and Institute of Information \& communications Technology Planning \& Evaluation (IITP) grant funded by the Korean government (MSIT) (RS-2019-II190079, Artificial Intelligence Graduate School Program (Korea University), RS-2024-00457882, AI Research Hub Project, and RS-2025-02653113, High-Performance Research AI Computing Infrastructure Support at the 2 PFLOPS Scale).

{
    \small
    \bibliographystyle{ieeenat_fullname}
    \bibliography{main}

\begin{thebibliography}{65}
\providecommand{\natexlab}[1]{#1}
\providecommand{\url}[1]{\texttt{#1}}
\expandafter\ifx\csname urlstyle\endcsname\relax
  \providecommand{\doi}[1]{doi: #1}\else
  \providecommand{\doi}{doi: \begingroup \urlstyle{rm}\Url}\fi

\bibitem[nvi(2023)]{nvidia2023a2f}
Omniverse audio2face: Generate expressive facial animation from just an audio source with nvidia’s deep learning ai technology.
\newblock \url{https://build.nvidia.com/nvidia/audio2face}, 2023.

\bibitem[Bao et~al.(2024)Bao, Zhang, Li, Zhang, Yang, Bao, Pollefeys, Zhang, and Cui]{bao2024geneavatar}
Chong Bao, Yinda Zhang, Yuan Li, Xiyu Zhang, Bangbang Yang, Hujun Bao, Marc Pollefeys, Guofeng Zhang, and Zhaopeng Cui.
\newblock Geneavatar: Generic expression-aware volumetric head avatar editing from a single image.
\newblock In \emph{Proceedings of the IEEE/CVF Conference on Computer Vision and Pattern Recognition}, pages 8952--8963, 2024.

\bibitem[Bescos et~al.(2018)Bescos, F{\'a}cil, Civera, and Neira]{bescos2018dynaslam}
Berta Bescos, Jos{\'e}~M F{\'a}cil, Javier Civera, and Jos{\'e} Neira.
\newblock Dynaslam: Tracking, mapping, and inpainting in dynamic scenes.
\newblock \emph{IEEE Robotics and Automation Letters}, 3\penalty0 (4):\penalty0 4076--4083, 2018.

\bibitem[Chai et~al.(2023)Chai, Zhang, He, Tan, Baltrusaitis, Wu, Li, Zhao, Yuan, and Bian]{chai2023hiface}
Zenghao Chai, Tianke Zhang, Tianyu He, Xu Tan, Tadas Baltrusaitis, HsiangTao Wu, Runnan Li, Sheng Zhao, Chun Yuan, and Jiang Bian.
\newblock Hiface: High-fidelity 3d face reconstruction by learning static and dynamic details.
\newblock In \emph{Proceedings of the IEEE/CVF International Conference on Computer Vision}, pages 9087--9098, 2023.

\bibitem[Chan et~al.(2022)Chan, Lin, Chan, Nagano, Pan, De~Mello, Gallo, Guibas, Tremblay, Khamis, et~al.]{chan2022efficient}
Eric~R Chan, Connor~Z Lin, Matthew~A Chan, Koki Nagano, Boxiao Pan, Shalini De~Mello, Orazio Gallo, Leonidas~J Guibas, Jonathan Tremblay, Sameh Khamis, et~al.
\newblock Efficient geometry-aware 3d generative adversarial networks.
\newblock In \emph{Proceedings of the IEEE/CVF conference on computer vision and pattern recognition}, pages 16123--16133, 2022.

\bibitem[Chen et~al.(2024)Chen, Wang, Li, Xiao, Zhang, Yao, and Liu]{chen2024monogaussianavatar}
Yufan Chen, Lizhen Wang, Qijing Li, Hongjiang Xiao, Shengping Zhang, Hongxun Yao, and Yebin Liu.
\newblock Monogaussianavatar: Monocular gaussian point-based head avatar.
\newblock In \emph{ACM SIGGRAPH 2024 Conference Papers}, pages 1--9, 2024.

\bibitem[Cho et~al.(2024)Cho, Lee, Yoon, Hong, Ko, Ahn, and Kim]{cho2024gaussiantalker}
Kyusun Cho, Joungbin Lee, Heeji Yoon, Yeobin Hong, Jaehoon Ko, Sangjun Ahn, and Seungryong Kim.
\newblock Gaussiantalker: Real-time high-fidelity talking head synthesis with audio-driven 3d gaussian splatting.
\newblock \emph{arXiv preprint arXiv:2404.16012}, 2024.

\bibitem[Chu and Harada(2024)]{chu2024generalizable}
Xuangeng Chu and Tatsuya Harada.
\newblock Generalizable and animatable gaussian head avatar.
\newblock \emph{Advances in Neural Information Processing Systems}, 37:\penalty0 57642--57670, 2024.

\bibitem[Cudeiro et~al.(2019)Cudeiro, Bolkart, Laidlaw, Ranjan, and Black]{cudeiro2019capture}
Daniel Cudeiro, Timo Bolkart, Cassidy Laidlaw, Anurag Ranjan, and Michael~J Black.
\newblock Capture, learning, and synthesis of 3d speaking styles.
\newblock In \emph{Proceedings of the IEEE/CVF conference on computer vision and pattern recognition}, pages 10101--10111, 2019.

\bibitem[Deng et~al.(2019)Deng, Guo, Niannan, and Zafeiriou]{deng2018arcface}
Jiankang Deng, Jia Guo, Xue Niannan, and Stefanos Zafeiriou.
\newblock Arcface: Additive angular margin loss for deep face recognition.
\newblock In \emph{CVPR}, 2019.

\bibitem[Dhamo et~al.(2025)Dhamo, Nie, Moreau, Song, Shaw, Zhou, and P{\'e}rez-Pellitero]{dhamo2025headgas}
Helisa Dhamo, Yinyu Nie, Arthur Moreau, Jifei Song, Richard Shaw, Yiren Zhou, and Eduardo P{\'e}rez-Pellitero.
\newblock Headgas: Real-time animatable head avatars via 3d gaussian splatting.
\newblock In \emph{European Conference on Computer Vision}, pages 459--476. Springer, 2025.

\bibitem[Dib et~al.(2021)Dib, Thebault, Ahn, Gosselin, Theobalt, and Chevallier]{dib2021towards}
Abdallah Dib, Cedric Thebault, Junghyun Ahn, Philippe-Henri Gosselin, Christian Theobalt, and Louis Chevallier.
\newblock Towards high fidelity monocular face reconstruction with rich reflectance using self-supervised learning and ray tracing.
\newblock In \emph{Proceedings of the IEEE/CVF International Conference on Computer Vision}, pages 12819--12829, 2021.

\bibitem[Fu et~al.(2024{\natexlab{a}})Fu, Liu, Kulkarni, Kautz, Efros, and Wang]{Fu_2024_CVPR}
Yang Fu, Sifei Liu, Amey Kulkarni, Jan Kautz, Alexei~A. Efros, and Xiaolong Wang.
\newblock Colmap-free 3d gaussian splatting.
\newblock In \emph{Proceedings of the IEEE/CVF Conference on Computer Vision and Pattern Recognition (CVPR)}, pages 20796--20805, 2024{\natexlab{a}}.

\bibitem[Fu et~al.(2024{\natexlab{b}})Fu, Liu, Kulkarni, Kautz, Efros, and Wang]{fu2024colmapfree}
Yang Fu, Sifei Liu, Amey Kulkarni, Jan Kautz, Alexei~A. Efros, and Xiaolong Wang.
\newblock Colmap-free 3d gaussian splatting.
\newblock In \emph{Proceedings of the IEEE/CVF Conference on Computer Vision and Pattern Recognition (CVPR)}, pages 20796--20805, 2024{\natexlab{b}}.

\bibitem[Gafni et~al.(2021)Gafni, Thies, Zollhofer, and Nie{\ss}ner]{gafni2021dynamic}
Guy Gafni, Justus Thies, Michael Zollhofer, and Matthias Nie{\ss}ner.
\newblock Dynamic neural radiance fields for monocular 4d facial avatar reconstruction.
\newblock In \emph{Proceedings of the IEEE/CVF Conference on Computer Vision and Pattern Recognition}, pages 8649--8658, 2021.

\bibitem[Gerig et~al.(2018)Gerig, Morel-Forster, Blumer, Egger, Luthi, Sch{\"o}nborn, and Vetter]{gerig2018morphable}
Thomas Gerig, Andreas Morel-Forster, Clemens Blumer, Bernhard Egger, Marcel Luthi, Sandro Sch{\"o}nborn, and Thomas Vetter.
\newblock Morphable face models-an open framework.
\newblock In \emph{2018 13th IEEE international conference on automatic face \& gesture recognition (FG 2018)}, pages 75--82. IEEE, 2018.

\bibitem[Giebenhain et~al.(2023)Giebenhain, Kirschstein, Georgopoulos, R{\"u}nz, Agapito, and Nie{\ss}ner]{giebenhain2023learning}
Simon Giebenhain, Tobias Kirschstein, Markos Georgopoulos, Martin R{\"u}nz, Lourdes Agapito, and Matthias Nie{\ss}ner.
\newblock Learning neural parametric head models.
\newblock In \emph{Proceedings of the IEEE/CVF Conference on Computer Vision and Pattern Recognition}, pages 21003--21012, 2023.

\bibitem[Grassal et~al.(2022)Grassal, Prinzler, Leistner, Rother, Nie{\ss}ner, and Thies]{grassal2022neural}
Philip-William Grassal, Malte Prinzler, Titus Leistner, Carsten Rother, Matthias Nie{\ss}ner, and Justus Thies.
\newblock Neural head avatars from monocular rgb videos.
\newblock In \emph{Proceedings of the IEEE/CVF Conference on Computer Vision and Pattern Recognition}, pages 18653--18664, 2022.

\bibitem[He et~al.(2024)He, Clausen, Ta{\c{s}}el, Ma, Pilarski, Xian, Rikker, Yu, Burgert, Yu, et~al.]{he2024diffrelight}
Mingming He, Pascal Clausen, Ahmet~Levent Ta{\c{s}}el, Li Ma, Oliver Pilarski, Wenqi Xian, Laszlo Rikker, Xueming Yu, Ryan Burgert, Ning Yu, et~al.
\newblock Diffrelight: Diffusion-based facial performance relighting.
\newblock \emph{arXiv preprint arXiv:2410.08188}, 2024.

\bibitem[Huang et~al.(2024)Huang, Yu, Chen, Geiger, and Gao]{huang20242d}
Binbin Huang, Zehao Yu, Anpei Chen, Andreas Geiger, and Shenghua Gao.
\newblock 2d gaussian splatting for geometrically accurate radiance fields.
\newblock In \emph{ACM SIGGRAPH 2024 Conference Papers}, pages 1--11, 2024.

\bibitem[Kemelmacher-Shlizerman and Basri(2010)]{kemelmacher20103d}
Ira Kemelmacher-Shlizerman and Ronen Basri.
\newblock 3d face reconstruction from a single image using a single reference face shape.
\newblock \emph{IEEE transactions on pattern analysis and machine intelligence}, 33\penalty0 (2):\penalty0 394--405, 2010.

\bibitem[Kerbl et~al.(2023)Kerbl, Kopanas, Leimk{\"u}hler, and Drettakis]{kerbl20233d}
Bernhard Kerbl, Georgios Kopanas, Thomas Leimk{\"u}hler, and George Drettakis.
\newblock 3d gaussian splatting for real-time radiance field rendering.
\newblock \emph{ACM Trans. Graph.}, 42\penalty0 (4):\penalty0 139--1, 2023.

\bibitem[Kirschstein et~al.(2023)Kirschstein, Qian, Giebenhain, Walter, and Nie{\ss}ner]{kirschstein2023nersemble}
Tobias Kirschstein, Shenhan Qian, Simon Giebenhain, Tim Walter, and Matthias Nie{\ss}ner.
\newblock Nersemble: Multi-view radiance field reconstruction of human heads.
\newblock \emph{ACM Transactions on Graphics (TOG)}, 42\penalty0 (4):\penalty0 1--14, 2023.

\bibitem[Kittler et~al.(2016)Kittler, Huber, Feng, Hu, and Christmas]{kittler20163d}
Josef Kittler, Patrik Huber, Zhen-Hua Feng, Guosheng Hu, and William Christmas.
\newblock 3d morphable face models and their applications.
\newblock In \emph{Articulated Motion and Deformable Objects: 9th International Conference, AMDO 2016, Palma de Mallorca, Spain, July 13-15, 2016, Proceedings 9}, pages 185--206. Springer, 2016.

\bibitem[Lee et~al.(2024)Lee, Lee, Sun, Ali, and Park]{lee2024deblurring}
Byeonghyeon Lee, Howoong Lee, Xiangyu Sun, Usman Ali, and Eunbyung Park.
\newblock Deblurring 3d gaussian splatting, 2024.

\bibitem[Lew et~al.(2025)Lew, Yoo, Kang, and Park]{lew2025towards}
Hah~Min Lew, Sahng-Min Yoo, Hyunwoo Kang, and Gyeong-Moon Park.
\newblock Towards high-fidelity head blending with chroma keying for industrial applications.
\newblock In \emph{2025 IEEE/CVF Winter Conference on Applications of Computer Vision (WACV)}, pages 6188--6196. IEEE, 2025.

\bibitem[Li et~al.(2024{\natexlab{a}})Li, Zhang, Bai, Zheng, Ning, Zhou, and Gu]{li2024talkinggaussian}
Jiahe Li, Jiawei Zhang, Xiao Bai, Jin Zheng, Xin Ning, Jun Zhou, and Lin Gu.
\newblock Talkinggaussian: Structure-persistent 3d talking head synthesis via gaussian splatting.
\newblock \emph{arXiv preprint arXiv:2404.15264}, 2024{\natexlab{a}}.

\bibitem[Li et~al.(2017)Li, Bolkart, Black, Li, and Romero]{li2017learning}
Tianye Li, Timo Bolkart, Michael~J Black, Hao Li, and Javier Romero.
\newblock Learning a model of facial shape and expression from 4d scans.
\newblock \emph{ACM Trans. Graph.}, 36\penalty0 (6):\penalty0 194--1, 2017.

\bibitem[Li et~al.(2024{\natexlab{b}})Li, De~Mello, Liu, Nagano, Iqbal, and Kautz]{li2024generalizable}
Xueting Li, Shalini De~Mello, Sifei Liu, Koki Nagano, Umar Iqbal, and Jan Kautz.
\newblock Generalizable one-shot 3d neural head avatar.
\newblock \emph{Advances in Neural Information Processing Systems}, 36, 2024{\natexlab{b}}.

\bibitem[Lin et~al.(2021)Lin, Ryabtsev, Sengupta, Curless, Seitz, and Kemelmacher-Shlizerman]{lin2021real}
Shanchuan Lin, Andrey Ryabtsev, Soumyadip Sengupta, Brian~L Curless, Steven~M Seitz, and Ira Kemelmacher-Shlizerman.
\newblock Real-time high-resolution background matting.
\newblock In \emph{Proceedings of the IEEE/CVF Conference on Computer Vision and Pattern Recognition}, pages 8762--8771, 2021.

\bibitem[Lin et~al.(2024)Lin, Li, Su, Zheng, Zhang, and Liu]{lin2024layga}
Siyou Lin, Zhe Li, Zhaoqi Su, Zerong Zheng, Hongwen Zhang, and Yebin Liu.
\newblock Layga: Layered gaussian avatars for animatable clothing transfer.
\newblock In \emph{ACM SIGGRAPH 2024 Conference Papers}, pages 1--11, 2024.

\bibitem[Loper et~al.(2023)Loper, Mahmood, Romero, Pons-Moll, and Black]{loper2023smpl}
Matthew Loper, Naureen Mahmood, Javier Romero, Gerard Pons-Moll, and Michael~J Black.
\newblock Smpl: A skinned multi-person linear model.
\newblock In \emph{Seminal Graphics Papers: Pushing the Boundaries, Volume 2}, pages 851--866. 2023.

\bibitem[Luo et~al.(2024)Luo, Ouyang, Zhao, Jiang, Zhang, Zhang, Yang, Xu, and Yu]{luo2024gaussianhair}
Haimin Luo, Min Ouyang, Zijun Zhao, Suyi Jiang, Longwen Zhang, Qixuan Zhang, Wei Yang, Lan Xu, and Jingyi Yu.
\newblock Gaussianhair: Hair modeling and rendering with light-aware gaussians.
\newblock \emph{arXiv preprint arXiv:2402.10483}, 2024.

\bibitem[Ma et~al.(2024)Ma, Weng, Shao, and Zhou]{ma20243d}
Shengjie Ma, Yanlin Weng, Tianjia Shao, and Kun Zhou.
\newblock 3d gaussian blendshapes for head avatar animation.
\newblock In \emph{ACM SIGGRAPH 2024 Conference Papers}, pages 1--10, 2024.

\bibitem[Mildenhall et~al.(2021)Mildenhall, Srinivasan, Tancik, Barron, Ramamoorthi, and Ng]{mildenhall2021nerf}
Ben Mildenhall, Pratul~P Srinivasan, Matthew Tancik, Jonathan~T Barron, Ravi Ramamoorthi, and Ren Ng.
\newblock Nerf: Representing scenes as neural radiance fields for view synthesis.
\newblock \emph{Communications of the ACM}, 65\penalty0 (1):\penalty0 99--106, 2021.

\bibitem[Morales et~al.(2021)Morales, Piella, and Sukno]{morales2021survey}
Araceli Morales, Gemma Piella, and Federico~M Sukno.
\newblock Survey on 3d face reconstruction from uncalibrated images.
\newblock \emph{Computer Science Review}, 40:\penalty0 100400, 2021.

\bibitem[M{\"u}ller et~al.(2022)M{\"u}ller, Evans, Schied, and Keller]{muller2022instant}
Thomas M{\"u}ller, Alex Evans, Christoph Schied, and Alexander Keller.
\newblock Instant neural graphics primitives with a multiresolution hash encoding.
\newblock \emph{ACM transactions on graphics (TOG)}, 41\penalty0 (4):\penalty0 1--15, 2022.

\bibitem[Palafox et~al.(2021)Palafox, Bo{\v{z}}i{\v{c}}, Thies, Nie{\ss}ner, and Dai]{palafox2021npms}
Pablo Palafox, Alja{\v{z}} Bo{\v{z}}i{\v{c}}, Justus Thies, Matthias Nie{\ss}ner, and Angela Dai.
\newblock Npms: Neural parametric models for 3d deformable shapes.
\newblock In \emph{Proceedings of the IEEE/CVF International Conference on Computer Vision}, pages 12695--12705, 2021.

\bibitem[Park et~al.(2021)Park, Sinha, Barron, Bouaziz, Goldman, Seitz, and Martin-Brualla]{park2021nerfies}
Keunhong Park, Utkarsh Sinha, Jonathan~T Barron, Sofien Bouaziz, Dan~B Goldman, Steven~M Seitz, and Ricardo Martin-Brualla.
\newblock Nerfies: Deformable neural radiance fields.
\newblock In \emph{Proceedings of the IEEE/CVF International Conference on Computer Vision}, pages 5865--5874, 2021.

\bibitem[Qian et~al.(2024)Qian, Kirschstein, Schoneveld, Davoli, Giebenhain, and Nie{\ss}ner]{qian2024gaussianavatars}
Shenhan Qian, Tobias Kirschstein, Liam Schoneveld, Davide Davoli, Simon Giebenhain, and Matthias Nie{\ss}ner.
\newblock Gaussianavatars: Photorealistic head avatars with rigged 3d gaussians.
\newblock In \emph{Proceedings of the IEEE/CVF Conference on Computer Vision and Pattern Recognition}, pages 20299--20309, 2024.

\bibitem[Rong et~al.(2024)Rong, Grigorev, Wang, Black, Thomaszewski, Tsalicoglou, and Hilliges]{rong2024gaussian}
Boxiang Rong, Artur Grigorev, Wenbo Wang, Michael~J Black, Bernhard Thomaszewski, Christina Tsalicoglou, and Otmar Hilliges.
\newblock Gaussian garments: Reconstructing simulation-ready clothing with photorealistic appearance from multi-view video.
\newblock \emph{arXiv preprint arXiv:2409.08189}, 2024.

\bibitem[Saputra et~al.(2018)Saputra, Markham, and Trigoni]{saputra2018visual}
Muhamad Risqi~U Saputra, Andrew Markham, and Niki Trigoni.
\newblock Visual slam and structure from motion in dynamic environments: A survey.
\newblock \emph{ACM Computing Surveys (CSUR)}, 51\penalty0 (2):\penalty0 1--36, 2018.

\bibitem[Schonberger and Frahm(2016)]{schonberger2016structure}
Johannes~L Schonberger and Jan-Michael Frahm.
\newblock Structure-from-motion revisited.
\newblock In \emph{Proceedings of the IEEE conference on computer vision and pattern recognition}, pages 4104--4113, 2016.

\bibitem[Shao et~al.(2024)Shao, Wang, Li, Wang, Lin, Zhang, Fan, and Wang]{shao2024splattingavatar}
Zhijing Shao, Zhaolong Wang, Zhuang Li, Duotun Wang, Xiangru Lin, Yu Zhang, Mingming Fan, and Zeyu Wang.
\newblock Splattingavatar: Realistic real-time human avatars with mesh-embedded gaussian splatting.
\newblock In \emph{Proceedings of the IEEE/CVF Conference on Computer Vision and Pattern Recognition}, pages 1606--1616, 2024.

\bibitem[Sun et~al.(2023)Sun, Wang, Wang, Li, Zhang, Zhang, and Liu]{sun2023next3d}
Jingxiang Sun, Xuan Wang, Lizhen Wang, Xiaoyu Li, Yong Zhang, Hongwen Zhang, and Yebin Liu.
\newblock Next3d: Generative neural texture rasterization for 3d-aware head avatars.
\newblock In \emph{Proceedings of the IEEE/CVF conference on computer vision and pattern recognition}, pages 20991--21002, 2023.

\bibitem[Sun et~al.(2024)Sun, Zhang, Wan, Zhou, Li, Ye, and Jiao]{sun2024correspondence}
Wei Sun, Xiaosong Zhang, Fang Wan, Yanzhao Zhou, Yuan Li, Qixiang Ye, and Jianbin Jiao.
\newblock Correspondence-guided sfm-free 3d gaussian splatting for nvs.
\newblock \emph{arXiv preprint arXiv:2408.08723}, 2024.

\bibitem[Ullman(1979)]{ullman1979interpretation}
Shimon Ullman.
\newblock The interpretation of structure from motion.
\newblock \emph{Proceedings of the Royal Society of London. Series B. Biological Sciences}, 203\penalty0 (1153):\penalty0 405--426, 1979.

\bibitem[Vaswani(2017)]{vaswani2017attention}
A Vaswani.
\newblock Attention is all you need.
\newblock \emph{Advances in Neural Information Processing Systems}, 2017.

\bibitem[Wang et~al.(2024)Wang, Kang, Sun, Qian, Wang, Bao, and Zhang]{wang2024mega}
Cong Wang, Di Kang, He-Yi Sun, Shen-Han Qian, Zi-Xuan Wang, Linchao Bao, and Song-Hai Zhang.
\newblock Mega: Hybrid mesh-gaussian head avatar for high-fidelity rendering and head editing.
\newblock \emph{arXiv preprint arXiv:2404.19026}, 2024.

\bibitem[Wang et~al.(2023)Wang, Xie, Li, Xu, Pun, and Gao]{wang2023gaussianhead}
Jie Wang, Jiu-Cheng Xie, Xianyan Li, Feng Xu, Chi-Man Pun, and Hao Gao.
\newblock Gaussianhead: Impressive head avatars with learnable gaussian diffusion.
\newblock \emph{arXiv preprint arXiv:2312.01632}, 2023.

\bibitem[Xiang et~al.(2023)Xiang, Gao, Guo, and Zhang]{xiang2023flashavatar}
Jun Xiang, Xuan Gao, Yudong Guo, and Juyong Zhang.
\newblock Flashavatar: High-fidelity digital avatar rendering at 300fps.
\newblock \emph{arXiv preprint arXiv:2312.02214}, 2023.

\bibitem[Xu et~al.(2023{\natexlab{a}})Xu, Song, Jiang, Zhang, Shi, Liu, Ma, Feng, and Luo]{xu2023omniavatar}
Hongyi Xu, Guoxian Song, Zihang Jiang, Jianfeng Zhang, Yichun Shi, Jing Liu, Wanchun Ma, Jiashi Feng, and Linjie Luo.
\newblock Omniavatar: Geometry-guided controllable 3d head synthesis.
\newblock In \emph{Proceedings of the IEEE/CVF Conference on Computer Vision and Pattern Recognition}, pages 12814--12824, 2023{\natexlab{a}}.

\bibitem[Xu et~al.(2023{\natexlab{b}})Xu, Wang, Zhao, Zhang, and Liu]{xu2023avatarmav}
Yuelang Xu, Lizhen Wang, Xiaochen Zhao, Hongwen Zhang, and Yebin Liu.
\newblock Avatarmav: Fast 3d head avatar reconstruction using motion-aware neural voxels.
\newblock In \emph{ACM SIGGRAPH 2023 Conference Proceedings}, pages 1--10, 2023{\natexlab{b}}.

\bibitem[Xu et~al.(2024)Xu, Chen, Li, Zhang, Wang, Zheng, and Liu]{xu2024gaussian}
Yuelang Xu, Benwang Chen, Zhe Li, Hongwen Zhang, Lizhen Wang, Zerong Zheng, and Yebin Liu.
\newblock Gaussian head avatar: Ultra high-fidelity head avatar via dynamic gaussians.
\newblock In \emph{Proceedings of the IEEE/CVF Conference on Computer Vision and Pattern Recognition}, pages 1931--1941, 2024.

\bibitem[Xuan et~al.(2024)Xuan, Li, Yao, Zhou, Sun, Chen, and Pan]{xuan2024faghead}
Yixin Xuan, Xinyang Li, Gongxin Yao, Shiwei Zhou, Donghui Sun, Xiaoxin Chen, and Yu Pan.
\newblock Faghead: Fully animate gaussian head from monocular videos.
\newblock \emph{arXiv preprint arXiv:2406.19070}, 2024.

\bibitem[Yu et~al.(2018)Yu, Wang, Peng, Gao, Yu, and Sang]{yu2018bisenet}
Changqian Yu, Jingbo Wang, Chao Peng, Changxin Gao, Gang Yu, and Nong Sang.
\newblock Bisenet: Bilateral segmentation network for real-time semantic segmentation.
\newblock In \emph{Proceedings of the European conference on computer vision (ECCV)}, pages 325--341, 2018.

\bibitem[Yuan et~al.(2024)Yuan, Li, Huang, De~Mello, Nagano, Kautz, and Iqbal]{yuan2024gavatar}
Ye Yuan, Xueting Li, Yangyi Huang, Shalini De~Mello, Koki Nagano, Jan Kautz, and Umar Iqbal.
\newblock Gavatar: Animatable 3d gaussian avatars with implicit mesh learning.
\newblock In \emph{Proceedings of the IEEE/CVF Conference on Computer Vision and Pattern Recognition}, pages 896--905, 2024.

\bibitem[Zhang et~al.(2018)Zhang, Isola, Efros, Shechtman, and Wang]{zhang2018unreasonable}
Richard Zhang, Phillip Isola, Alexei~A Efros, Eli Shechtman, and Oliver Wang.
\newblock The unreasonable effectiveness of deep features as a perceptual metric.
\newblock In \emph{Proceedings of the IEEE conference on computer vision and pattern recognition}, pages 586--595, 2018.

\bibitem[Zhao et~al.(2024)Zhao, Bao, Li, Qiu, and Liu]{zhao2024psavatar}
Zhongyuan Zhao, Zhenyu Bao, Qing Li, Guoping Qiu, and Kanglin Liu.
\newblock Psavatar: A point-based morphable shape model for real-time head avatar creation with 3d gaussian splatting.
\newblock \emph{arXiv preprint arXiv:2401.12900}, 2024.

\bibitem[Zheng et~al.(2024)Zheng, Wen, Li, Zhang, Su, Chang, Zhao, Lv, Zhang, Zhang, et~al.]{zheng2024headgap}
Xiaozheng Zheng, Chao Wen, Zhaohu Li, Weiyi Zhang, Zhuo Su, Xu Chang, Yang Zhao, Zheng Lv, Xiaoyuan Zhang, Yongjie Zhang, et~al.
\newblock Headgap: Few-shot 3d head avatar via generalizable gaussian priors.
\newblock \emph{arXiv preprint arXiv:2408.06019}, 2024.

\bibitem[Zheng et~al.(2022)Zheng, Abrevaya, B{\"u}hler, Chen, Black, and Hilliges]{zheng2022avatar}
Yufeng Zheng, Victoria~Fern{\'a}ndez Abrevaya, Marcel~C B{\"u}hler, Xu Chen, Michael~J Black, and Otmar Hilliges.
\newblock Im avatar: Implicit morphable head avatars from videos.
\newblock In \emph{Proceedings of the IEEE/CVF Conference on Computer Vision and Pattern Recognition}, pages 13545--13555, 2022.

\bibitem[Zheng et~al.(2023)Zheng, Yifan, Wetzstein, Black, and Hilliges]{zheng2023pointavatar}
Yufeng Zheng, Wang Yifan, Gordon Wetzstein, Michael~J Black, and Otmar Hilliges.
\newblock Pointavatar: Deformable point-based head avatars from videos.
\newblock In \emph{Proceedings of the IEEE/CVF conference on computer vision and pattern recognition}, pages 21057--21067, 2023.

\bibitem[Zhou et~al.(2024)Zhou, Ma, Fan, and Yang]{zhou2024headstudio}
Zhenglin Zhou, Fan Ma, Hehe Fan, and Yi Yang.
\newblock Headstudio: Text to animatable head avatars with 3d gaussian splatting.
\newblock \emph{arXiv preprint arXiv:2402.06149}, 2024.

\bibitem[Zielonka et~al.(2022)Zielonka, Bolkart, and Thies]{zielonka2022towards}
Wojciech Zielonka, Timo Bolkart, and Justus Thies.
\newblock Towards metrical reconstruction of human faces.
\newblock In \emph{European conference on computer vision}, pages 250--269. Springer, 2022.

\bibitem[Zielonka et~al.(2023)Zielonka, Bolkart, and Thies]{zielonka2023instant}
Wojciech Zielonka, Timo Bolkart, and Justus Thies.
\newblock Instant volumetric head avatars.
\newblock In \emph{Proceedings of the IEEE/CVF Conference on Computer Vision and Pattern Recognition}, pages 4574--4584, 2023.

\end{thebibliography}
}

\clearpage

\section{Appendix}

In the Appendix, we show additional extensive experimental results.
First, we show details of the quantitative comparison of both datasets: DynamicFace and SplattingAvatar \cite{shao2024splattingavatar}.
Second, we depict the qualitative results on various self- and cross-reenactment and novel-view synthesis scenarios.
Then, we conduct additional ablation studies on APS, FLAME mouth modification, and isolating the contribution of each component.
Followed by the visualization of APS during mouth deformation, we also demonstrate training and inference details and the preprocessing of our model.
Then, we elucidate the details of the mesh modification process, dataset, and baselines.
Finally, we discuss the broader impacts of GeoAvatar, along with two examples: interactable digital human and virtual presentation.

\subsection{Additional Quantitative Comparison}
\label{app:quantitative}
In Table \ref{tab:app_ours} and Table \ref{tab:app_splat}, we compared the self-reenactment results of each video in SplattingAvatar \cite{shao2024splattingavatar} and DynamicFace, respectively.
For a fair and thorough comparison, we utilized every 10 subjects in each video dataset.
We denote the name of each subject with the quantitative results in Table \ref{tab:app_ours} and Table \ref{tab:app_splat}.
In both datasets, our model shows the best performance in almost every video.
Specifically, ours shows the best results in 19 videos out of 20 videos.

Moreover, we demonstrated quantitative comparisons with other models for a more comprehensive evaluation as shown in Table \ref{tab:app_cross}. 
For evaluation metrics, we measured a cosine similarity by employing off-the-shelf models: 1) insightface\footnote[1]{\href{https://github.com/deepinsight/insightface}{https://github.com/deepinsight/insightface}}~\cite{deng2018arcface} for ``ID preservation'' and 2) Facial-Expression-Recognition model\footnote[1]{\href{https://github.com/WuJie1010/Facial-Expression-Recognition.Pytorch}{https://github.com/WuJie1010/Facial-Expression-Recognition.Pytorch}} trained on FER2013 for ``Expression''.
Still, ours shows the superior results on both ID preservation and expression scores.

\begin{table}[h]
\resizebox{\columnwidth}{!}{%
\begin{tabular}{c|cccccc}
\toprule
& INSTA & 3DGS & SplattingAvatar & FlashAvatar & GaussianAvatars & \textbf{Ours} \\
\midrule
ID preservation $\uparrow$ & 0.850 & 0.783 & 0.873 & \underline{0.889} & 0.831 & \textbf{0.906} \\
Expression $\uparrow$ & 0.712 & 0.458 & 0.667 & \underline{0.717} & 0.681 & \textbf{0.750} \\
\bottomrule
\end{tabular}%
}
\caption{Quantitative evaluation on cross-reenactment.}
\label{tab:app_cross}
\end{table}

\subsection{Additional Qualitative Comparison}
\label{app:experiments}

\begin{figure}[!h]
\centering
\includegraphics[width=\columnwidth]{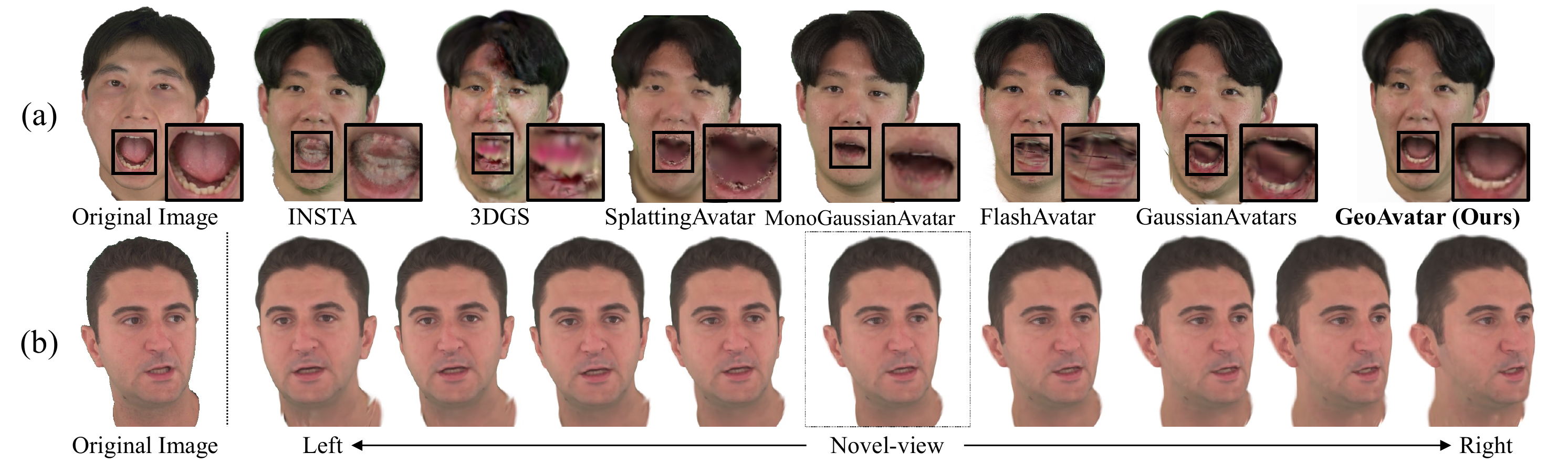}
\caption{
Cross-reenactment and novel-view synthesis.
}
\label{fig:appendix_exp_novel}
\end{figure}

\noindent In Figure \ref{fig:appendix_exp_novel}(a), we compare cross-reenactments by utilizing the EMO-1-shout+laugh sequence of id061 in NeRSemble for challenging expressions.
Ours shows stable results, while the baselines suffer from severe artifacts.
In Figure \ref{fig:appendix_exp_novel}(b), we show novel-view synthesis under the same identity and expression.
Ours successfully generates high-quality images while maintaining consistency in identity.

In Figure \ref{fig:appendix_monocular}, we show extensive self-reenactment results of ours, compared to baselines.
Indeed, ours shows notably better performance on the mouth region, \eg, handling artifacts on the second row, and resolution of teeth in the fourth and sixth row.
Moreover, our model shows robust and high-resolution outputs on other facial regions, \eg, the iris in the first and the fifth rows, and the ear in the third row.

In Figure \ref{fig:appendix_cross}, we show extensive cross-reenactment results of ours, compared to baselines.
To evaluate the robustness of each model more thoroughly, we utilize the source and target actors from different datasets, \eg, the source from SplattingAvatar when the target is from DynamicFace, and vice versa.
Even in this harsh scenario, ours shows consistently robust reenactment results, while preserving the identity of the source actor and mimicking the expression of the target actor well.
On the other hand, other models suffer from severe artifacts, occurring by notable distribution differences between the source and the target actors.

In Figure \ref{fig:appendix_novel}, we show extensive novel-view synthesis results of ours, compared to baselines.
Ours shows consistently robust results on various actors.
Especially, we emphasize that in the case when the original image has the extreme facial degree, \eg, the fourth row in the Figure \ref{fig:appendix_novel}, ours can generate the face robustly in more extreme viewpoints.
In contrast, other models suffer from artifacts, \eg, INSTA, 3DGS, and GaussianAvatars, spiking artifacts, \eg, FlashAvatar, ghosting effects, \eg, INSTA.
Please check our \href{https://hahminlew.github.io/geoavatar/}{project page} for more visualization results.

\subsection{Additional Ablation Results}
\label{app:abl}

\begin{figure}[h]
\begin{center}
\includegraphics[width=1\columnwidth]{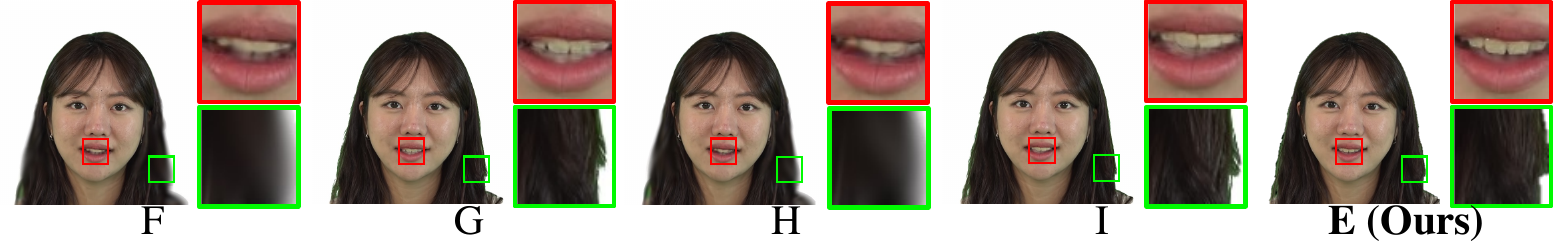}
\end{center}
\caption{
\textbf{Additional qualitative ablation results.}
We show qualitative results for additional ablation models for thorough evaluations.
Models that only utilizes the rigid set,\ie, F and H, show notably blurred results for the region where the FLAME mesh cannot cover the ground truth geometry, \eg, hair.
On the other hand, setting every region to flexible set, \ie, G and I, introduces undesirable artifacts for the mouth animation.
}
\label{fig:appendix_abl}
\end{figure}
\begin{table}[h]
    \centering
        \resizebox{\columnwidth}{!}{
        \begin{tabular}{ccccccc}
            \toprule[1.1pt]
            Configs & Sets & \makecell{FLAME mouth} & \makecell{MSE \scriptsize($10^{-3}$)} $\downarrow$ & PSNR $\uparrow$ & SSIM $\uparrow$ & \makecell{LPIPS \scriptsize($10^{-1}$)} $\downarrow$ \\ 
            \midrule
            F & Rigid & & 0.960 & 30.291 & 0.923 & 0.975 \\
            G & Flexible & & 0.991 & 30.356 & 0.919 & 0.653 \\
            H & Rigid & \checkmark & 0.901 & 30.471 & 0.923 & 0.840 \\
            I & Flexible & \checkmark & \underline{0.853} & \underline{31.318} & \underline{0.930} & \underline{0.629} \\ 
            \midrule
            \textbf{E (Ours)} & APS & \checkmark & \textbf{0.748} & \textbf{32.697} & \textbf{0.942} & \textbf{0.513} \\
            \bottomrule[1.1pt]
        \end{tabular}%
        }
        \caption{
        \textbf{Additional quantitative ablation results.}
        We show quantitative results for additional ablation models for thorough evaluations.
        While applying either APS or FLAME mouth slightly improve the performance, they show the synergestic effect when applied together.
        }
        \label{tab:app_ablations}
\end{table}

\begin{table}[t]
\centering
\resizebox{0.9\columnwidth}{!}{%
\begin{tabular}{llcccc}
    \toprule[1.1pt]
    & Configuration & \makecell{MSE \scriptsize($10^{-3}$)} $\downarrow$ & PSNR $\uparrow$ & SSIM $\uparrow$ & \makecell{LPIPS \scriptsize($10^{-1}$)} $\downarrow$ \\ 
    \midrule
    A & \textbf{Ours} w/o APS & 0.853 & 31.318 & 0.930 & 0.629 \\
    B & \textbf{Ours} w/o FLAME mouth & 0.912 & 30.551 & 0.929 & 0.561 \\
    C & \textbf{Ours} w/o Part-wise deformation & 0.815 & 32.162 & 0.941 & 0.540 \\
    D & \textbf{Ours} w/o $\mathcal{L}_{angle}$ & \textbf{0.733} & \textbf{32.751} & \underline{0.941} & \underline{0.519} \\
    \midrule
    E & \textbf{Ours} & \underline{0.748} & \underline{32.697} & \textbf{0.942} & \textbf{0.513} \\
    \bottomrule[1.1pt]
\end{tabular}
}
\caption{Quantitative ablation results by isolating the contribution.}
\label{tab:ablation_re}
\end{table}

\noindent In the extension from Table \ref{tab:ablation}, we add more settings for thorough ablations.
First, to evaluate the effectiveness of training initialization strategy, we train the baseline model with assuming every Gaussian as the rigid set, \ie, F, and the flexible set, \ie, G.
Since the original GaussianAvatars \cite{qian2024gaussianavatars} utilizes threshold for position loss same with the threshold of the flexible set, A and G is originally same.
Then, we apply FLAME mouth modification, \ie, mesh modification and part-wise deformation, to each model, \ie, H and I, respectively.
Table \ref{tab:app_ablations} shows the average quantitative results of each model, including our final model, \ie, E.
We utilize every model in SplattingAvatar and DynamicFace to obtain the average result in Table \ref{tab:app_ablations}.

First, in F and G, both model shows inferior results in both qualitative and quantitative ways.
In specific, as shown in Figure \ref{app:abl}, F shows severe artifacts in the hair region, which needs high flexibility during training.
By applying FLAME modification and deformation to F and G, \ie, H and I, respectively, improves the performance, it still shows artifacts in hair or mouth.

Then, in H and I only applies FLAME mouth modification and part-wise deformation, without applying APS.
Though only applying these improves both quantitative and qualitative results notably, still it shows worse result than our final model.
We argue that since APS helps model to train each part flexibly, it is helpful to improve the overall quality of every part, either rigid or flexible.
Indeed, our final model, \ie, E, shows the best result on both quantitative and qualitative ways.

To better clarify the contribution of each module, we also perform ablations by isolating each module, as shown in Table \ref{tab:ablation_re}.
Especially, in Config C, \ie, excluding part-wise deformation based on MLP, shows a notable degradation compared to the original model, \ie, Config E, which clearly shows the effectiveness of the MLP deformation.

\subsection{Visualization of APS and mouth deformation}
\label{app:vis}

\begin{figure}[h]
\begin{center}
\includegraphics[width=1\columnwidth]{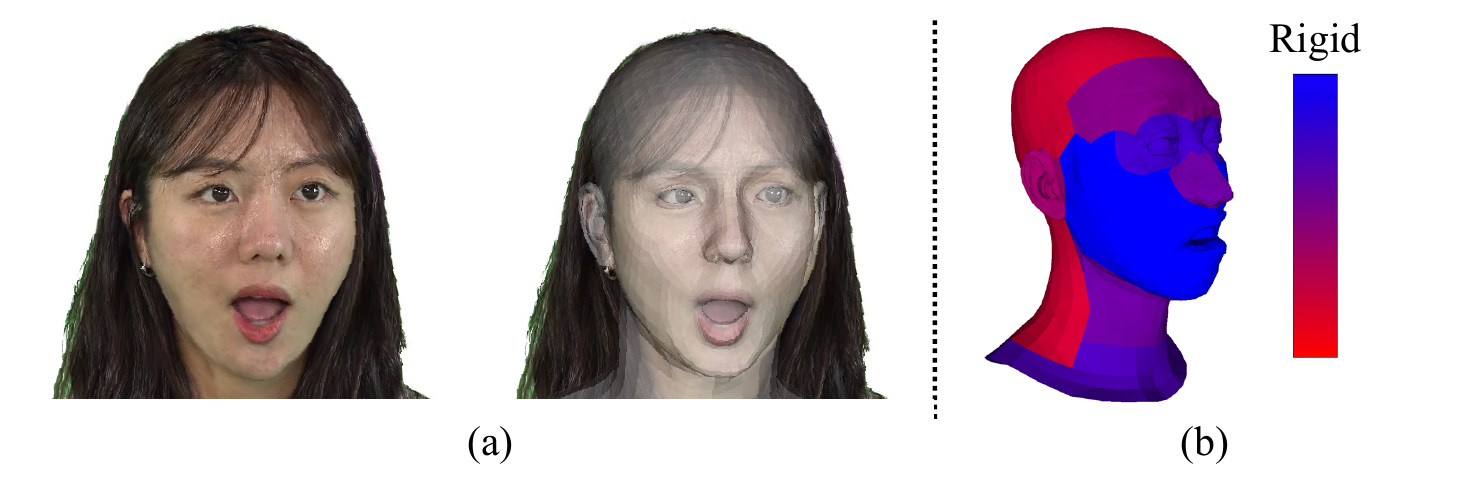}
\end{center}
\caption{
(a) Rendered results with the mesh,
(b) Distribution of rigid and flexible sets.
}
\label{fig:appendix_mesh}
\end{figure}

We visualized the fitted mesh during mouth deformation in Figure~\ref{fig:appendix_mesh}(a).
We also plot the distributions of rigid and flexible sets in Figure \ref{fig:appendix_mesh}(b).
We indeed check that the distribution follows the human intuition, \eg, rigid for facial regions, flexible for scalp and neck. 
Several regions, \eg, forehead and ears, show varying results among subjects, \ie, denoted as purple region, which justifies the dynamic allocation of rigid and flexible sets of APS.

\subsection{Training and Inference Details}
\label{app:training}

\begin{table}[h]
\resizebox{\columnwidth}{!}{%
\begin{tabular}{cccccc}
\toprule
Efficiency & FlashAvatar & GaussianAvatars & MonoGaussianAvatar & GaussianHeadAvatar & \textbf{Ours} \\
\midrule
\makecell{Time (hrs)} $\downarrow$ & \textbf{1.66} & 9.25 & 7.90 & 19.92 & \underline{4.90} \\
\makecell{Speed (FPS)} $\uparrow$ & \textbf{291.20} & 19.11 & 5.91 & 6.25 & \underline{71.52} \\
\bottomrule

\end{tabular}%
}
\caption{Efficiency comparisons on training time (hrs) and inference speed (FPS).}
\label{app:efficiency}
\end{table}

\begin{table}[h]
\resizebox{\columnwidth}{!}{%
\begin{tabular}{cccccccc}
\toprule
Steps ($10^3$) & 50 & 100 & APS & 150 & 200 & 250 & 300 \\
\midrule
SplattingAvatar & 33.95 & 42.22 & - & 34.67 & 43.61 & 46.12 & 47.63 \\
DynamicFace & 35.17 & 44.86 & - & 36.48 & 47.80 & 51.03 & 51.42 \\
\bottomrule

\end{tabular}%
}
\caption{Gaussian numbers ($10^3$) on each dataset.}
\label{app:gaunum}
\end{table}

\noindent In Table \ref{app:efficiency}, we compare training and inference speeds with a single RTX 3090.
Ours shows enough inference efficiency, \ie, $>60$ FPS, for real-time scenario.
We show numbers of Gaussians at training and inference (300,000 steps) stages in Table \ref{app:gaunum}.

\subsection{Preprocessing}
\label{app:preprocessing}
We can coarsely divide preprocessing into two steps; masking and FLAME tracking. In the following paragraph, we explain the details.

\begin{figure}[h]
\begin{center}
\includegraphics[width=1\columnwidth]{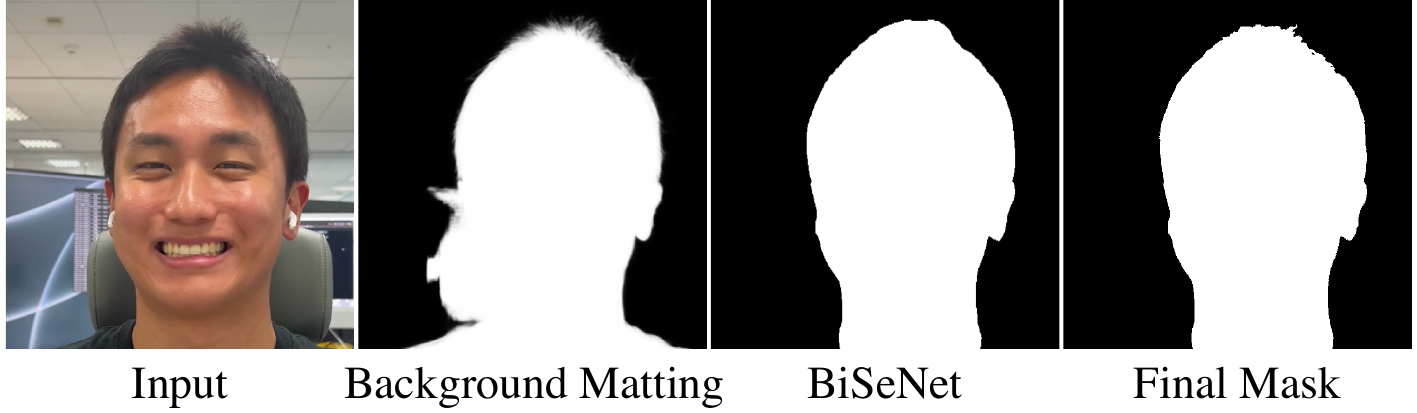}
\end{center}
\caption{
\textbf{Comparison of each mask.}
Background Matting yields a noisy mask, \ie, containing non-human parts, while BiSeNet yields an over-smoothed mask.
We intersect two masks and obtain the final mask for training.
}
\label{fig:mask}
\end{figure}

\noindent\textbf{Masking.} Since we target to generate human avatars, we have to distinguish human parts from the others, \eg, clothes and background, as other baselines \cite{zielonka2023instant, xiang2023flashavatar, shao2024splattingavatar} did.
We conjugate two masking logics, Background Matting \cite{lin2021real} and BiSeNet \cite{yu2018bisenet}.
Though Background Matting can distinguish the foreground objects from the background, it still contains non-human parts, \eg, chairs that the subject is sitting in.
On the other hand, BiSeNet can distinguish human parts and non-human parts accurately, but it returns over-smoothed masks.
Consequently, by utilizing the overlapped region of each mask, we can obtain an accurate and sharp mask for the human part.
Figure \ref{fig:mask} shows the visualization of each mask.
We apply the aforementioned mask logic for preprocessing DynamicFace and SplattingAvatar datasets, but not for the NeRSemble dataset since they offer the preprocessed mask together with the images.

\noindent\textbf{FLAME Tracking.} We utilize the modification of MICA \cite{zielonka2022towards}, which utilizes a pre-trained model that returns FLAME shape parameters from a single image, and an additional image-wise FLAME tracking model.
We modified the original MICA to work on the 2023 version of FLAME to utilize its revised eye region mesh, which originally worked on the 2020 version.
Moreover, we optimize the FLAME neck, global rotation, and translation parameters, which are excluded in the original MICA, along with other FLAME parameters, \eg, expression, jaw, and eye pose, for better tracking results.

\noindent\textbf{Hyperparameters.}
\label{app:hyperparameters}
We set the hyperparameters for the training as in Table \ref{tab:hyperparameter}.
Except for the hyperparameters mentioned, we follow the settings of GaussianAvatars \cite{qian2024gaussianavatars}.
\begin{table}[h]
\centering
\begin{tabular}{c|l|c}
\toprule
Symbol & Parameter Description & Value \\
\midrule
$n$ & number of FLAME parts & 10 \\
$\tau_{r}$ & threshold for the rigid set & 0.1 \\
$\tau_{f}$ & threshold for the flexible set & 2.0 \\
$\tau_{\varphi}$ & threshold for $\mathcal{L}_{angle}$ & 0.78 \\
$\lambda$ & weight for D-SSIM & 0.2 \\
$N$ & iterations before APS & 100000 \\
- & total iteration & 200000 \\
\bottomrule
\end{tabular}
\caption{\textbf{Hyperparameter settings of GeoAvatar.}
We utilize the hyperparameters mentioned above for training GeoAvatar.}
\label{tab:hyperparameter}
\end{table}

\subsection{Structural Details}
\label{app:structure}

\begin{figure}[h]
\begin{center}
\includegraphics[width=1\columnwidth]{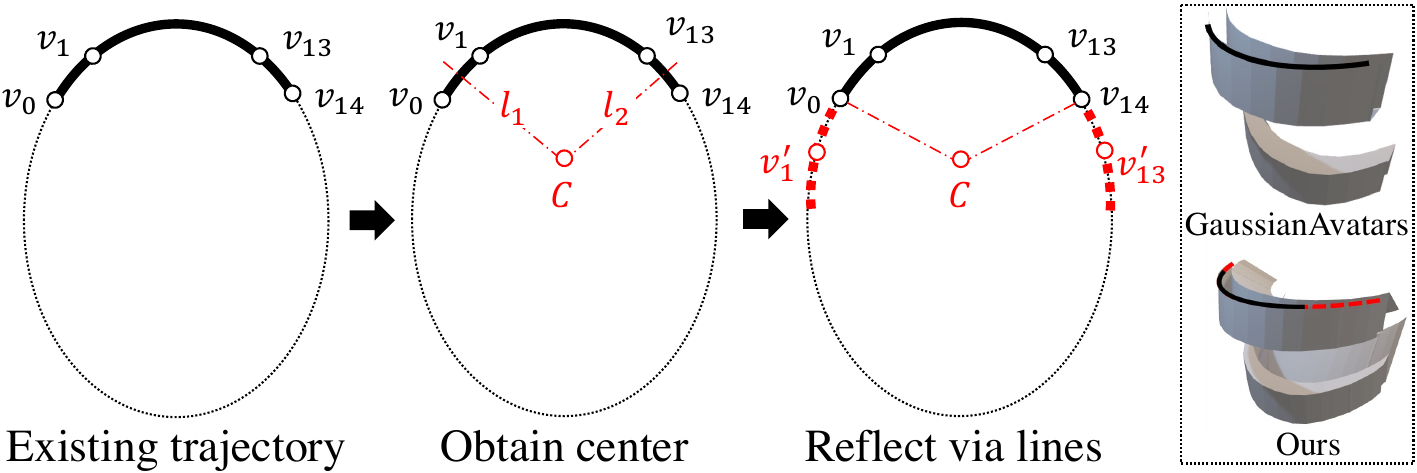}
\end{center}
\caption{
\textbf{Teeth trajectory extension.}
Assuming that the existing teeth trajectory forms an arc of the circle, we calculate the pseudo-center $C$ of the circle by using the perpendicular bisector line of $\overline{v_0v_1}$ and $\overline{v_{13}v_{14}}$.
Then, we reflect the existing vertices of the teeth by either $\overline{Cv_{0}}$ or $\overline{Cv_{14}}$, to extend the trajectory smoothly.}
\label{fig:appendix_teeth}
\end{figure}

\noindent{\textbf{Mesh Modification.}} Though FLAME \cite{li2017learning} covers various expressions and joint movements of the face, the absence of mouth interior structures deteriorates the expressiveness of teeth and mouth interior \cite{xiang2023flashavatar, li2024talkinggaussian, qian2024gaussianavatars}.
Consequently, GaussianAvatars \cite{qian2024gaussianavatars} adds teeth by duplicating the vertices trajectory of lip rings of FLAME.
As shown in Figure \ref{fig:appendix_teeth}, the mesh corresponding to frontal teeth can be generated.
However, this cannot represent the geometry of molar teeth and mouth interior structure, \eg, palate or tongue.

For better representation, we incorporate molar teeth and mouth interiors into the FLAME structure.
First, we generate the frontal teeth by utilizing the vertex trajectory of the lip rings \cite{qian2024gaussianavatars}.
Empirically, we observe that the teeth vertices lie on the $xz$-plane, \ie, they share the same $y$-axis value, and their trajectory approximates the shape of an ellipsoid.
Given that the curvature of the teeth trajectory is relatively small, we hypothesize that it can be approximated as an arc of a circle.
Due to the symmetry of a circle about its center, the arc can be extended smoothly by reflecting it across the center.
We apply this approach to the pseudo-arc trajectory of the teeth.
To this end, we identify the pseudo-center of the circle and extend the teeth trajectory to generate the molar teeth, as illustrated in Figure \ref{fig:appendix_teeth}.

First, we utilize the two leftmost vertices, \ie, $v_{0}=(x_{0},y_{0},z_{0})$ and $v_{1}=(x_1,y_{0},z_1)$, and the two rightmost vertices, \ie, $v_{13}=(x_{13},y_{0},z_{13})$ and $v_{14}=(x_{14},y_{0},z_{14})$, out of 15 vertices that constructs the teeth trajectory ring.
First, to obtain the pseudo-center, we obtain the intersection point $C$ between the perpendicular bisector of $\overline{v_{0}v_{1}}$, \ie, ${l_{1}}$, and the perpendicular bisector of $\overline{v_{13}v_{14}}$, \ie, ${l_{2}}$.
Each perpendicular bisector can be obtained as follows:
\begin{flalign*}
    & l_{1}: z - \frac{z_{1}+z_{2}}{2} = -\frac{x_{2}-x_{1}}{z_{2}-z_{1}}(x-\frac{x_1+x_2}{2}), \\
    & l_{2}: z - \frac{z_{13}+z_{14}}{2}= -\frac{x_{14}-x_{13}}{z_{14}-z_{13}}(x-\frac{x_{13}+x_{14}}{2}).
\end{flalign*}

Then, we reflect the 5 vertices located on the left side, \ie, $v_{i\in\{1,\cdots 5\}}$, with the line $\overline{Cv_{0}}$.
In the same way, we reflect the 5 vertices located on the right side, \ie, $v_{i\in\{9,\cdots 13\}}$, with the line $\overline{Cv_{14}}$.
Finally, we can obtain the new teeth trajectory which includes the molar teeth, denoted as the red dot line in Figure \ref{fig:appendix_teeth}.
After generating the trajectory of the teeth, we shift it backward to generate vertices for the palate and the mouth floor.

\subsection{Dataset Details}
\label{app:dataset}

\begin{table}[!h]
\resizebox{\columnwidth}{!}{%
\begin{tabular}{c|ccccc}
    \toprule[1.1pt]
    Dataset & \#ID & \#Expressions & Resolution & Total Time (min) & Disk Space (GB) \\
    \midrule
    NerFace (CVPR 2021) & 3 & 1 & 1920$\times$1080 & 6.44 & 3.79 \\ 
    IMAvatar (CVPR 2022) & 4 & 11 & 512$\times$512 & 7.08 & 3.39 \\
    \textbf{DynamicFace} & \textbf{10} & \textbf{20}\tnote{*} & \textbf{3840}$\times$\textbf{2160} & \textbf{32.25} & \textbf{18.92} \\
    \bottomrule
\end{tabular}
}
\caption{Comparison of monocular video-based datasets.}
\label{tab:dataset_re}
\end{table}

\noindent Our proposed dataset, \textbf{DynamicFace} is designed to capture a wide range of facial movements, enabling the generation of avatars capable of dynamic motion.
DynamicFace consists of 10 videos, each recording a single actor performing various facial expressions provided by the instruction.
During recording, actors are instructed to shake their heads slowly to record the various facial degrees with a single camera.
Nine subjects are recorded by a single Sony AX700 camcorder with a chromakey background.
The remaining subject is recorded by a single iPhone14 with a normal background.
In Figure \ref{fig:appendix_dynamicface}, we show the sample frames of DynamicFace.
We also compare the details of monocular video-based dataset features in Table \ref{tab:dataset_re}.

\subsection{Baseline Descriptions}
\label{app:baseline}
\noindent{\textbf{INSTA.}}
The instant volumetric head avatars (INSTA) framework embeds a dynamic neural radiance field into a surface-aligned multi-resolution grid around a 3D parametric face model. It employs a deformation field guided by FLAME to map points between deformed and canonical spaces and uses 3DMM-driven geometry regularization for depth alignment. The approach utilizes neural graphics primitives with multi-resolution hash encoding to represent the radiance field, enabling reconstruction and rendering based on monocular RGB videos.

\noindent{\textbf{3D Gaussian Splatting.}}
Unlike an existing 3D representation module\cite{mildenhall2021nerf, muller2022instant} which implicitly encodes the color and density information of the volume inside MLP, 3D Gaussian Splatting (3DGS) explicitly represents the 3D volume using the mean and 3D variance of Gaussian distribution.
Moreover, the efficient tile-based rasterizer of 3DGS enables remarkably faster rendering than the existing module.
However, 3DGS requires accurate pre-computed camera poses \cite{sun2024correspondence, Fu_2024_CVPR}, \eg, obtained by COLMAP \cite{schonberger2016structure}.
Moreover, primitive 3DGS can be applied only to static scenes, which is definitely not appropriate for dynamic avatar generation.
To adapt its property in the avatar generation, we utilize FLAME meshes instead of COLMAP to initialize the Gaussian points.

\noindent{\textbf{SplattingAvatar.}}
SplattingAvatar proposes a binding strategy between Gaussian and FLAME mesh, which forces Gaussians to move together with FLAME mesh, which is deformed by FLAME coefficients.
Since SplattingAvatar does not utilize additional regularization terms to locate Gaussians nearby bonded triangles, it utilizes the walking triangles strategy to adaptively change the bonded triangle of Gaussian.

\noindent{\textbf{MonoGaussianAvatar.}}
MonoGaussianAvatar proposes a point-based 3D Gaussian head avatar generation framework.
They enhance the point insertion and deletion strategy which prunes away invisible points via thresholding of opacity.
They also utilizes the Gaussian deformation field to preserve the accessories.

\noindent{\textbf{FlashAvatar.}}
FlashAvatar also utilizes the strategy of binding Gaussians to the FLAME mesh by using its UV map.
Then they utilize Gaussian offset models to deform Gaussians by the animation of FLAME meshes.
To enhance the mouth interior generation performance, FlashAvatar adds additional faces to fill the mouth cavity, using vertices on the lip.
Moreover, they utilize the masked loss which focuses on the mouth region.

\noindent{\textbf{GaussianAvatars.}}
GaussianAvatars proposes a multi-view-based Gaussian head avatar creation framework by rigging 3D Gaussian splats to 3DMM faces. This framework employs adaptive density control with binding inheritance, ensuring that newly created or pruned 3D Gaussian splats remain consistently attached to their parent triangles on the FLAME mesh during densification.

In GaussianAvatars, the linear blend skinning weight of upper teeth is rigged to the neck, \ie, head movements, while that of lower teeth is rigged to the jaw.
Consequently, upper and lower teeth movements are determined by depending on head and jaw movements, respectively.
To mitigate this, we propose a deformation network that offers offsets part-wisely, to translate each part independently from the FLAME parameters using the offsets, which is unavailable in GaussianAvatars.

\subsection{Additional Comparison Study on a One-shot-based Method}

\begin{table}[h]
\centering
\resizebox{\columnwidth}{!}{%
\begin{tabular}{lllccccc}
    \toprule[1.1pt]
    Dataset & Setting & Method & \makecell{MSE \scriptsize($10^{-3}$)} $\downarrow$ & PSNR $\uparrow$ & SSIM $\uparrow$ & \makecell{LPIPS \scriptsize($10^{-1}$)} $\downarrow$ & ID preservation $\uparrow$ \\
    \midrule
    \multirow{3}{*}{SplattingAvatar} & One-shot & GAGAvatar~\cite{chu2024generalizable} & 5.108 & 23.541 & 0.875 & 1.121 & 0.832 \\
    \cmidrule{2-8}
    & Video & \textbf{Ours} & \textbf{0.884} & \textbf{32.635} & \textbf{0.965} & \textbf{0.367} & \textbf{0.944} \\
    \midrule
    \multirow{3}{*}{DynamicFace} & One-shot & GAGAvatar~\cite{chu2024generalizable} & 2.421 & 26.406 & 0.850 & 1.485 & 0.827 \\
    \cmidrule{2-8}
    & Video & \textbf{Ours} & \textbf{0.612} & \textbf{32.760} & \textbf{0.919} & \textbf{0.660} & \textbf{0.931} \\
    \bottomrule[1.1pt]
\end{tabular}
}
\caption{Quantitative comparisons on one-shot and video settings.}
\label{tab:app_oneshot}
\end{table}
\begin{figure}[!h]
\centering
\includegraphics[width=\columnwidth]{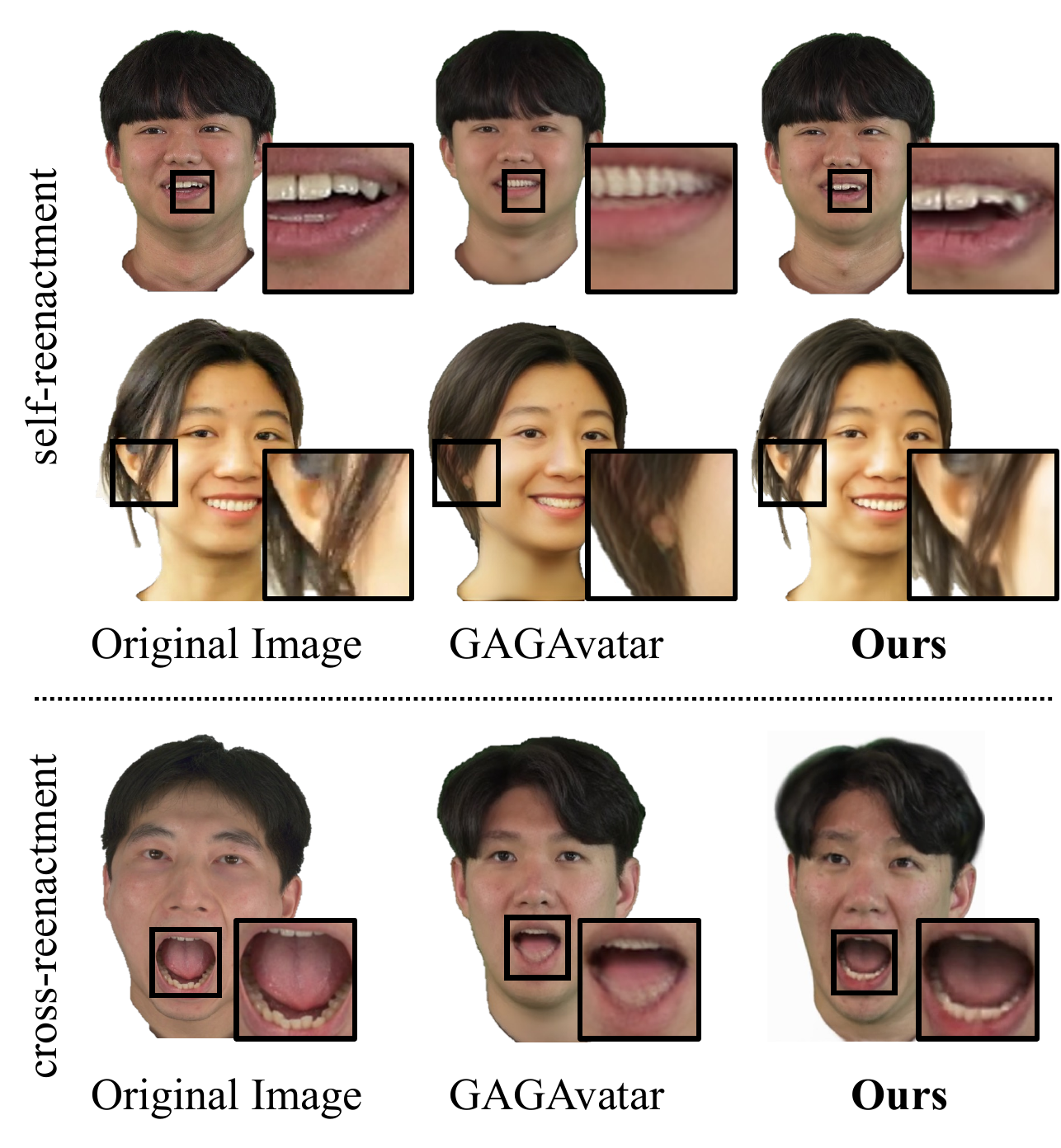}
\caption{
Comparisons on a one-shot-based baseline (GAGAvatar~\cite{chu2024generalizable}) via self- and cross-reenactment.
}
\label{fig:appendix_oneshot}
\end{figure}

\noindent Our work focuses on a single video-based setting for preserving identity under expressive variations, \ie, high-quality personalization.
Unlike one-shot or few-shot-based methods that risk entangling identity, video sequences provide sufficient intra-subject variation and temporal consistency to robustly learn identity-specific structures such as the mouth interior.
Furthermore, image-based methods often require large-scale pretraining on external datasets, whereas our method achieves high-quality results without any pretraining, making it practical for real-world applications.
In Table~\ref{tab:app_oneshot} and Figure~\ref{fig:appendix_oneshot}, we compared GeoAvatar (video) and state-of-the-art image-based avatar generation method, GAGAvatar~\cite{chu2024generalizable} (one-shot).
Ours achieves significantly higher identity preservation, while GAGAvatar struggles to preserve identity.

\subsection{Broader Impacts}
The proposed GeoAvatar framework demonstrates remarkable versatility, offering a broad spectrum of potential applications across diverse industries. By leveraging its capability to generate realistic, speech-driven 3D avatars, GeoAvatar has the potential to significantly enhance user experiences in domains such as entertainment, education, customer service, and healthcare. Its modular design allows seamless integration with complementary technologies, including large language models (LLMs), text-to-speech (TTS) models, and advanced 3D animation techniques, positioning it as a robust solution for creating engaging, interactive digital human experiences.

The following sections present two examples of industrial-like applications built on GeoAvatar as shown in Figure~\ref{fig:application}. Notably, these demonstrations were developed using training data captured with a single smartphone, emphasizing the scalability and accessibility of our method.

\subsection{Application \#1: Interactable Digital Human}
To highlight the capabilities of GeoAvatar, we integrated a large language model (LLM), a text-to-speech (TTS) model, and speech-driven 3D facial animation modules from the NVIDIA Audio2Face framework~\cite{nvidia2023a2f}. This configuration enabled the development of a real-time interactable digital human demo capable of engaging in natural, dynamic conversations with users.

In this system, the LLM processes textual user inputs and generates contextually appropriate responses. 
These responses are then converted into natural-sounding speech by the TTS model. Finally, the Audio2Face module drives a 3D avatar, synchronizing facial expressions, lip movements, and head gestures to match the speech output. 
This seamless integration results in an immersive, life-like digital human experience, illustrating GeoAvatar’s potential for real-time applications in virtual environments, customer engagement, and interactive storytelling. The demo video is in the \href{https://hahminlew.github.io/geoavatar}{project page}. We utilized CHANGER~\cite{lew2025towards} for seamless head blending into the original scenes.

\subsection{Application \#2: Virtual Presentation}
Expanding on recent advancements in speech-driven animation, we incorporated the state-of-the-art speech-to-facial animation module (S2F) into the GeoAvatar framework. This integration facilitated the generation of highly realistic talking head avatars characterized by natural head movements, expressive facial animations, and precise lip synchronization.

The use of S2F allows GeoAvatar to produce avatars with distinct personalities and speaking styles, making them suitable for a variety of applications, such as virtual presentations, personalized virtual assistants, and digital influencers. The module’s ability to capture nuanced head motions and deliver high-quality lip-syncing enhances the realism and engagement of these avatars, particularly in use cases where authentic communication and emotional expressiveness are critical. This example underscores the framework’s flexibility and its capacity to deliver diverse, high-fidelity digital human experiences.

\begin{table*}[t]
\centering\footnotesize
\begin{tabular}{l|ccccccc}
\toprule
Model & INSTA & 3DGS & SplattingAvatar & MonoGaussianAvatar & FlashAvatar & GaussianAvatars & Ours \\
\bottomrule
\textbf{Subject} & \multicolumn{7}{c}{subject\_001} \\
\midrule
MSE \scriptsize($10^{-3}$) $\downarrow$
& 0.741
& 0.987
& \underline{0.666}
& 0.987
& 1.112 
& 0.776 
& \textbf{0.350} \\
PSNR $\uparrow$ 
& 31.402 
& 30.127 
& \underline{31.817} 
& 30.154
& 29.646 
& 31.191 
& \textbf{34.670} \\
SSIM $\uparrow$ & \underline{0.909} & 0.896 & 0.894 & 0.900 & 0.890 & 0.894 & \textbf{0.929} \\
LPIPS \scriptsize($10^{-1}$) $\downarrow$ & 0.984 & 1.437 & 1.270 & 1.447 & \underline{0.695} & 0.728 & \textbf{0.633} \\
\bottomrule
\textbf{Subject} & \multicolumn{7}{c}{subject\_002} \\
\midrule
MSE \scriptsize($10^{-3}$) $\downarrow$ & 1.689 & 1.857 & 1.772 & 2.544 & 2.160 & \underline{1.497} & \textbf{0.787} \\
PSNR $\uparrow$ & 27.746 & 27.351 & 27.531 & 25.974 & 26.716 & \underline{28.284} & \textbf{31.063} \\
SSIM $\uparrow$ & 0.820 & 0.814 & 0.795 & \underline{0.821} & 0.814 & 0.792 & \textbf{0.874} \\
LPIPS \scriptsize($10^{-1}$) $\downarrow$ & 1.723 & 2.020 & 1.728 & 1.479 & \underline{0.881} & 0.982 & \textbf{0.775} \\
\bottomrule
\textbf{Subject} & \multicolumn{7}{c}{subject\_003} \\
\midrule
MSE \scriptsize($10^{-3}$) $\downarrow$ & \underline{0.553} & 0.757 & 1.036 & 1.217 & 1.256 & 0.865 & \textbf{0.310} \\
PSNR $\uparrow$ & \underline{32.617} & 31.294 & 29.930 & 29.340 & 29.088 & 30.701 & \textbf{35.155} \\
SSIM $\uparrow$ & \underline{0.918} & 0.904 & 0.884 & 0.891 & 0.888 & 0.905 & \textbf{0.940} \\
LPIPS \scriptsize($10^{-1}$) $\downarrow$ & 0.905 & 1.408 & 1.215 & 1.286 & 0.715 & \underline{0.626} & \textbf{0.518} \\
\bottomrule

\textbf{Subject} & \multicolumn{7}{c}{subject\_004} \\
\midrule
MSE \scriptsize($10^{-3}$) $\downarrow$ & 2.173 & 1.680 & 1.328 & 1.330 & 2.548 & \underline{1.080} & \textbf{0.507} \\
PSNR $\uparrow$ & 27.039 & 28.147 & 28.975 & 28.866 & 25.957 & \underline{29.733} & \textbf{33.192} \\
SSIM $\uparrow$ & 0.880 & 0.865 & 0.871 & \underline{0.893} & 0.869 & 0.868 & \textbf{0.914} \\
LPIPS \scriptsize($10^{-1}$) $\downarrow$ & 1.355 & 1.704 & 1.160 & 1.131 & 0.817 & \underline{0.627} & \textbf{0.547} \\
\bottomrule

\textbf{Subject} & \multicolumn{7}{c}{subject\_005} \\
\midrule
MSE \scriptsize($10^{-3}$) $\downarrow$ & 1.283 & 1.128 & 1.109 & 1.380 & 1.219 & \underline{1.062} & \textbf{0.440} \\
PSNR $\uparrow$ & 29.017 & 29.566 & 29.650 & 28.715 & 29.244 & \underline{29.826} & \textbf{33.703} \\
SSIM $\uparrow$ & 0.882 & 0.867 & 0.865 & \underline{0.874} & 0.865 & 0.869 & \textbf{0.913} \\
LPIPS \scriptsize($10^{-1}$) $\downarrow$ & 1.444 & 2.029 & 1.328 & 1.464 & \underline{0.679} & 0.893 & \textbf{0.643} \\
\bottomrule

\textbf{Subject} & \multicolumn{7}{c}{subject\_006} \\
\midrule
MSE \scriptsize($10^{-3}$) $\downarrow$ & 2.064 & 1.649 & 2.008 & 2.390 & 1.892 & \underline{1.091} & \textbf{0.688} \\
PSNR $\uparrow$ & 27.115 & 27.995 & 27.056 & 26.268 & 27.350 & \underline{29.688} & \textbf{31.673} \\
SSIM $\uparrow$ & 0.879 & 0.864 & 0.855 & \underline{0.880} & 0.861 & 0.872 & \textbf{0.916} \\
LPIPS \scriptsize($10^{-1}$) $\downarrow$ & 1.618 & 2.215 & 1.686 & 1.527 & \underline{0.803} & 0.850 & \textbf{0.702} \\
\bottomrule

\textbf{Subject} & \multicolumn{7}{c}{subject\_007} \\
\midrule
MSE \scriptsize($10^{-3}$) $\downarrow$ & 1.756 & 0.915 & 0.994 & 1.346 & 1.573 & \underline{0.840} & \textbf{0.374} \\
PSNR $\uparrow$ & 27.608 & 30.518 & 30.092 & 28.735 & 28.092 & \underline{30.843} & \textbf{34.357} \\
SSIM $\uparrow$ & 0.912 & 0.905 & 0.897 & 0.905 & 0.895 & \underline{0.903} & \textbf{0.939} \\
LPIPS \scriptsize($10^{-1}$) $\downarrow$ & 1.147 & 1.342 & 1.251 & 1.233 & 0.680 & \underline{0.536} & \textbf{0.522} \\
\bottomrule

\textbf{Subject} & \multicolumn{7}{c}{subject\_008} \\
\midrule
MSE \scriptsize($10^{-3}$) $\downarrow$ & \underline{2.124} & 2.995 & 2.322 & 2.166 & 3.127 & 3.030 & \textbf{1.564} \\
PSNR $\uparrow$ & \underline{26.994} & 25.368 & 26.435 & 26.710 & 25.287 & 25.288 & \textbf{28.198} \\
SSIM $\uparrow$ & \underline{0.863} & 0.844 & 0.852 & 0.871 & 0.857 & 0.844 & \textbf{0.900} \\
LPIPS \scriptsize($10^{-1}$) $\downarrow$ & 1.563 & 2.003 & 1.653 & 1.455 & \textbf{0.910} & 1.269 & \underline{1.020} \\
\bottomrule

\textbf{Subject} & \multicolumn{7}{c}{subject\_009} \\
\midrule
MSE \scriptsize($10^{-3}$) $\downarrow$ & 2.119 & 2.952 & 1.957 & 1.324 & \underline{1.735} & 2.008 & \textbf{0.666} \\
PSNR $\uparrow$ & 27.029 & 25.444 & 27.162 & \underline{28.892} & 27.965 & 27.023 & \textbf{31.835} \\
SSIM $\uparrow$ & \underline{0.925} & 0.910 & 0.901 & 0.923 & 0.917 & 0.908 & \textbf{0.940} \\
LPIPS \scriptsize($10^{-1}$) $\downarrow$ & 0.944 & 1.132 & 1.200 & 1.005 & \textbf{0.601} & 0.810 & \underline{0.629} \\
\bottomrule

\textbf{Subject} & \multicolumn{7}{c}{subject\_010} \\
\midrule
MSE \scriptsize($10^{-3}$) $\downarrow$ & 0.946 & 1.113 & 1.065 & 1.494 & 1.489 & \textbf{0.427} & \underline{0.435} \\
PSNR $\uparrow$ & 30.310 & 29.657 & 29.780 & 28.302 & 28.383 & \textbf{33.840} & \underline{33.753} \\
SSIM $\uparrow$ & 0.887 & 0.877 & 0.876 & 0.898 & 0.880 & \textbf{0.933} & \underline{0.924} \\
LPIPS \scriptsize($10^{-1}$) $\downarrow$ & 1.204 & 1.702 & 1.287 & 1.189 & 0.666 & \textbf{0.595} & \underline{0.614} \\
\bottomrule

\end{tabular}
\caption{\textbf{Additional quantitative results.}
Detailed quantitative comparison results of each subject from DynamicFace.
For a fair comparison, we utilized all 10 subjects without omission.
\textbf{Bold} indicates the best and \underline{underline} indicates the second.
}\label{tab:app_ours}
\end{table*}

\clearpage
\begin{table*}[t]
\centering\footnotesize
\begin{tabular}{l|ccccccc}
\toprule
Model & INSTA & 3DGS & SplattingAvatar & MonoGaussianAvatar & FlashAvatar & GaussianAvatars & Ours \\
\bottomrule
\textbf{Subject} & \multicolumn{7}{c}{bala} \\
\midrule
MSE \scriptsize($10^{-3}$) $\downarrow$ & 1.396 & 1.2691 & 4.428 & 0.968 & \underline{0.576} & 0.661 & \textbf{0.485} \\
PSNR $\uparrow$ & 28.609 & 29.010 & 23.550 & 30.426 & \underline{32.510} & 31.822 & \textbf{33.193} \\
SSIM $\uparrow$ & 0.936 & 0.914 & 0.922 & 0.928 & 0.938 & \underline{0.957} & \textbf{0.968} \\
LPIPS \scriptsize($10^{-1}$) $\downarrow$ & 0.816 & 1.424 & 0.715 & 0.819 & \underline{0.346} & 0.350 & \textbf{0.320} \\
\bottomrule
\textbf{Subject} & \multicolumn{7}{c}{biden} \\
\midrule
MSE \scriptsize($10^{-3}$) $\downarrow$ & 0.661 & 0.804 & 1.898 & 0.749 & \underline{0.713} & 0.783 & \textbf{0.291} \\
PSNR $\uparrow$ & \underline{31.994} & 31.205 & 27.308 & 31.356 & 31.531 & 31.303 & \textbf{35.586} \\
SSIM $\uparrow$ & \underline{0.969} & 0.958 & 0.955 & 0.959 & \underline{0.969} & 0.956 & \textbf{0.982} \\
LPIPS \scriptsize($10^{-1}$) $\downarrow$ & 0.359 & 0.452 & 0.412 & 0.433 & \underline{0.239} & 0.239 & \textbf{0.169} \\
\bottomrule

\textbf{Subject} & \multicolumn{7}{c}{malte\_1} \\
\midrule
MSE \scriptsize($10^{-3}$) $\downarrow$ & 1.273 & 1.644 & 2.779 & 0.926 & 1.033 & \underline{0.559} & \textbf{0.387} \\
PSNR $\uparrow$ & 29.186 & 28.076 & 25.631 & 30.669 & 30.418 & \underline{32.665} & \textbf{34.224} \\
SSIM $\uparrow$ & 0.946 & 0.933 & 0.937 & 0.944 & 0.949 & \underline{0.970} & \textbf{0.976} \\
LPIPS \scriptsize($10^{-1}$) $\downarrow$ & 0.579 & 0.807 & 0.509 & 0.561 & 0.343 & \underline{0.305} & \textbf{0.273} \\
\bottomrule

\textbf{Subject} & \multicolumn{7}{c}{marcel} \\
\midrule
MSE \scriptsize($10^{-3}$) $\downarrow$ & \underline{1.180} & 2.794 & 2.228 & 1.089 & 1.850 & 2.902 & \textbf{1.149} \\
PSNR $\uparrow$ & \underline{29.635} & 25.807 & 26.497 & 29.791 & 27.404 & 24.457 & \textbf{29.639} \\
SSIM $\uparrow$ & 0.947 & 0.929 & 0.940 & \underline{0.948} & 0.931 & 0.913 & \textbf{0.960} \\
LPIPS \scriptsize($10^{-1}$) $\downarrow$ & \underline{0.556} & 1.075 & \textbf{0.530} & 0.591 & 0.634 & 0.706 & 0.588 \\
\bottomrule

\textbf{Subject} & \multicolumn{7}{c}{nf\_01} \\
\midrule
MSE \scriptsize($10^{-3}$) $\downarrow$ & 1.705 & 1.904 & 3.643 & \underline{1.505} & 1.644 & 1.961 & \textbf{1.054} \\
PSNR $\uparrow$ & 27.891 & 27.470 & 24.458 & \underline{28.484} & 28.114 & 27.399 & \textbf{30.027} \\
SSIM $\uparrow$ & 0.945 & 0.933 & 0.935 & 0.939 & \underline{0.951} & 0.931 & \textbf{0.966} \\
LPIPS \scriptsize($10^{-1}$) $\downarrow$ & 0.677 & 1.030 & 0.631 & 0.694 & \underline{0.473} & 0.722 & \textbf{0.431} \\
\bottomrule

\textbf{Subject} & \multicolumn{7}{c}{nf\_03} \\
\midrule
MSE \scriptsize($10^{-3}$) $\downarrow$ & 1.171 & 1.962 & 2.001 & 1.576 & 1.383 & \underline{0.636} & \textbf{0.492} \\
PSNR $\uparrow$ & 29.637 & 27.249 & 27.063 & 28.226 & 28.857 & \underline{30.403} & \textbf{33.286} \\
SSIM $\uparrow$ & 0.941 & 0.923 & 0.934 & 0.935 & 0.941 & \underline{0.943} & \textbf{0.968} \\
LPIPS \scriptsize($10^{-1}$) $\downarrow$ & 0.565 & 0.940 & 0.465 & 0.596 & 0.392 & \underline{0.360} & \textbf{0.353} \\
\bottomrule

\textbf{Subject} & \multicolumn{7}{c}{obama} \\
\midrule
MSE \scriptsize($10^{-3}$) $\downarrow$ & 3.264 & 1.546 & 3.245 & 1.045 & 1.371 & \underline{0.854} & \textbf{0.234} \\
PSNR $\uparrow$ & 27.231 & 29.317 & 25.073 & 30.212 & \underline{30.387} & 29.243 & \textbf{36.801} \\
SSIM $\uparrow$ & 0.951 & 0.937 & 0.945 & 0.958 & \underline{0.963} & 0.935 & \textbf{0.981} \\
LPIPS \scriptsize($10^{-1}$) $\downarrow$ & 0.487 & 0.565 & 0.558 & 0.427 & 0.335 & \underline{0.267} & \textbf{0.157} \\
\bottomrule

\textbf{Subject} & \multicolumn{7}{c}{person\_0004} \\
\midrule
MSE \scriptsize($10^{-3}$) $\downarrow$ & 6.881 & 6.726 & 3.478 & 6.078 & 9.676 & \textbf{2.579} & \underline{2.308} \\
PSNR $\uparrow$ & 24.045 & 24.536 & 24.648 & 22.345 & 24.503 & \underline{26.377} & \textbf{30.349} \\
SSIM $\uparrow$ & 0.909 & 0.904 & 0.937 & 0.913 & 0.928 & \underline{0.937} & \textbf{0.947} \\
LPIPS \scriptsize($10^{-1}$) $\downarrow$ & 1.102 & 1.365 & 0.697 & 1.803 & 0.704 & \underline{0.632} & \textbf{0.431} \\
\bottomrule

\textbf{Subject} & \multicolumn{7}{c}{wojtek\_1} \\
\midrule
MSE \scriptsize($10^{-3}$) $\downarrow$ & 1.203 & 0.738 & 3.422 & 0.645 & 0.423 & \underline{0.364} & \textbf{0.282} \\
PSNR $\uparrow$ & 29.269 & 31.534 & 24.683 & 32.119 & 33.867 & \underline{34.454} & \textbf{35.535} \\
SSIM $\uparrow$ & 0.957 & 0.949 & 0.938 & 0.956 & 0.964 & \underline{0.979} & \textbf{0.981} \\
LPIPS \scriptsize($10^{-1}$) $\downarrow$ & 0.563 & 0.630 & 0.541 & 0.521 & 0.234 & \underline{0.220} & \textbf{0.217} \\
\bottomrule

\textbf{Subject} & \multicolumn{7}{c}{yufeng} \\
\midrule
MSE \scriptsize($10^{-3}$) $\downarrow$ & 5.559 & 8.312 & 4.145 & 4.373 & \underline{3.727} & 5.497 & \textbf{2.061} \\
PSNR $\uparrow$ & 23.331 & 21.227 & 24.412 & 24.501 & \underline{25.463} & 23.122 & \textbf{27.708} \\
SSIM $\uparrow$ & 0.878 & 0.849 & 0.888 & 0.893 & \underline{0.900} & 0.860 & \textbf{0.924} \\
LPIPS \scriptsize($10^{-1}$) $\downarrow$ & 1.084 & 1.896 & 0.823 & 0.885 & \underline{0.741} & 1.334 & \textbf{0.729} \\
\bottomrule

\end{tabular}
\caption{\textbf{Additional quantitative results.}
Detailed quantitative comparison results of each subject provided by SplattingAvatar \cite{shao2024splattingavatar}.
For a fair comparison, we utilized all 10 subjects without omission.
\textbf{Bold} indicates the best and \underline{underline} indicates the second.
}\label{tab:app_splat}
\end{table*}

\clearpage


\begin{figure*}[t]
\begin{center}
\includegraphics[width=1\textwidth]{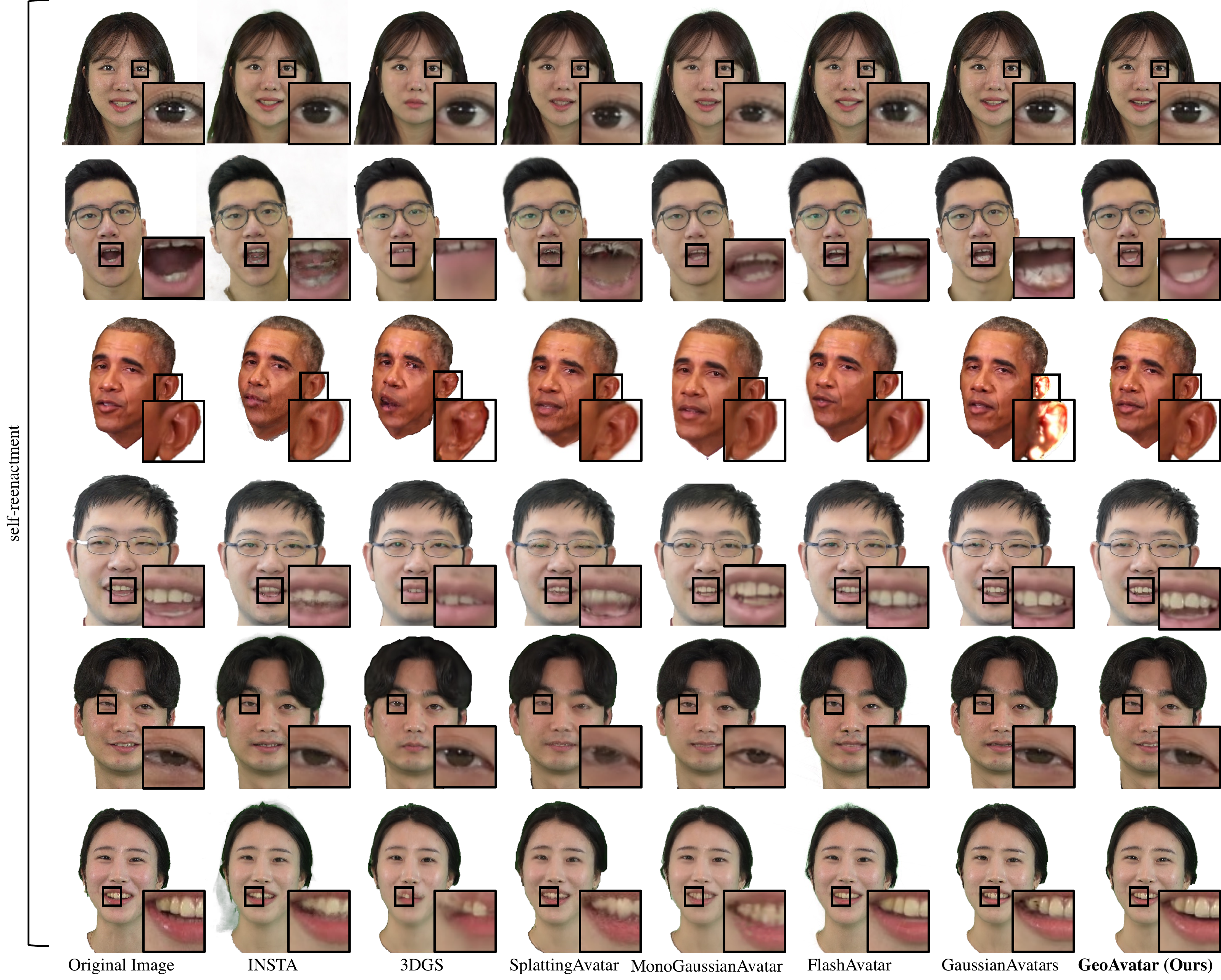}
\end{center}
\caption{
\textbf{Additional self-reenactment synthesis results.}
We show additional self-reenactment results of ours, compared to baselines, on SplattingAvatar and DynamicFace datasets.
Our shows high-resolution results not only on regions where FLAME geometry is accurate, \eg, eyes, but also on the regions where FLAME geometry is erroneous or absent, \eg, ears and mouth interior.
However, baselines struggle to generate high-resolution results, \eg, the first and fifth rows, or to prevent artifacts, \eg, the second, fourth, and sixth rows.
}
\label{fig:appendix_monocular}
\end{figure*}
\begin{figure*}[t]
\begin{center}
\includegraphics[width=1\textwidth]{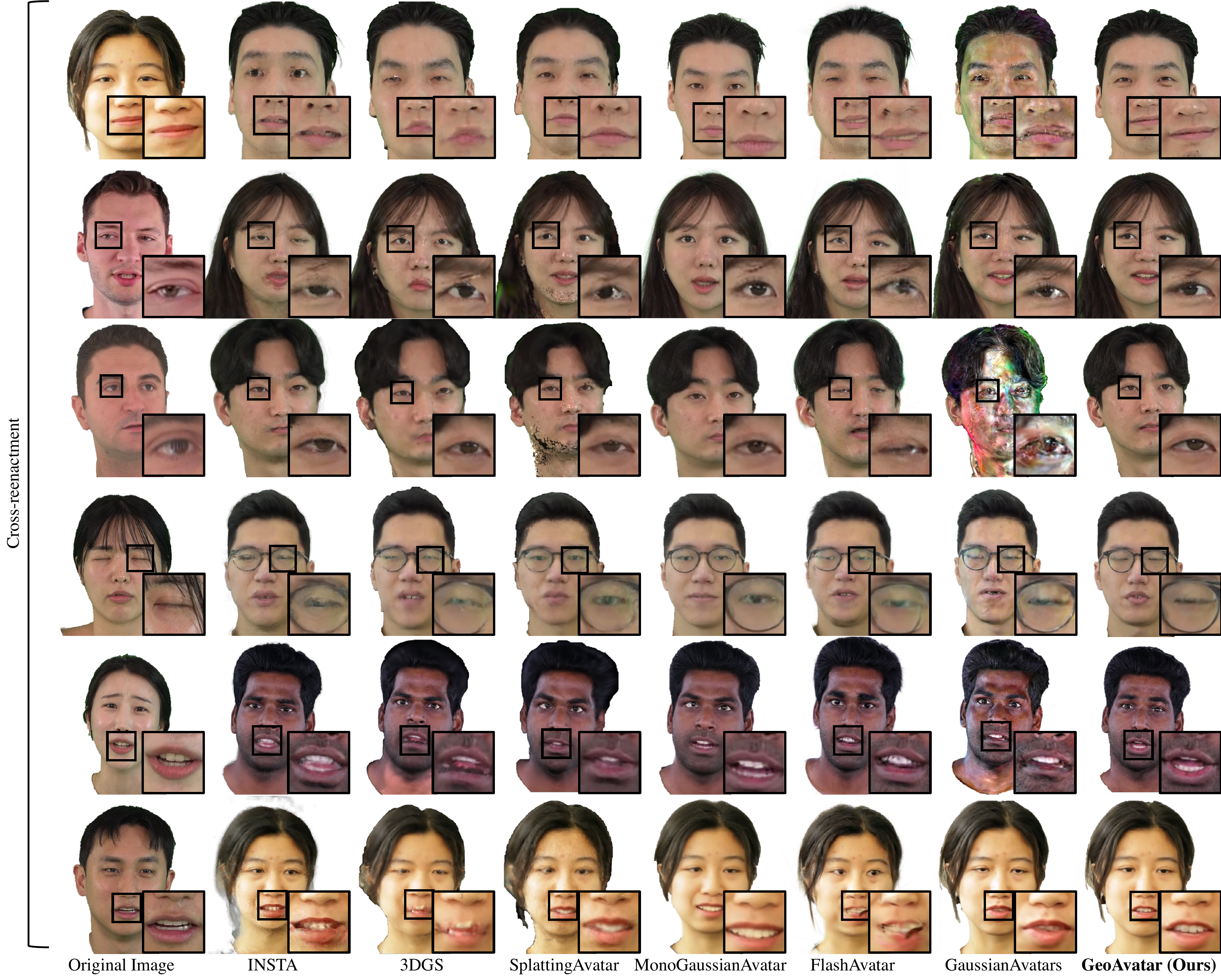}
\end{center}
\caption{
\textbf{Additional cross-reenactment synthesis results.}
We show additional cross-reenactment results of ours, compared to baselines, on SplattingAvatar and DynamicFace datasets.
To evaluate models thoroughly, we utilize the source and target actors from different datasets, \eg, SplattinAvatar and DynamicFace.
Ours shows robust generation results without artifacts, while baselines suffer from severe artifacts.
In specific, ours shows outstanding results for rendering eye regions, \eg, the first, second, and third rows, including accessories, \eg, eyeglasses in the third row.
Best viewed zoom-in.}
\label{fig:appendix_cross}
\end{figure*}
\begin{figure*}[t]
\begin{center}
\includegraphics[width=1\textwidth]{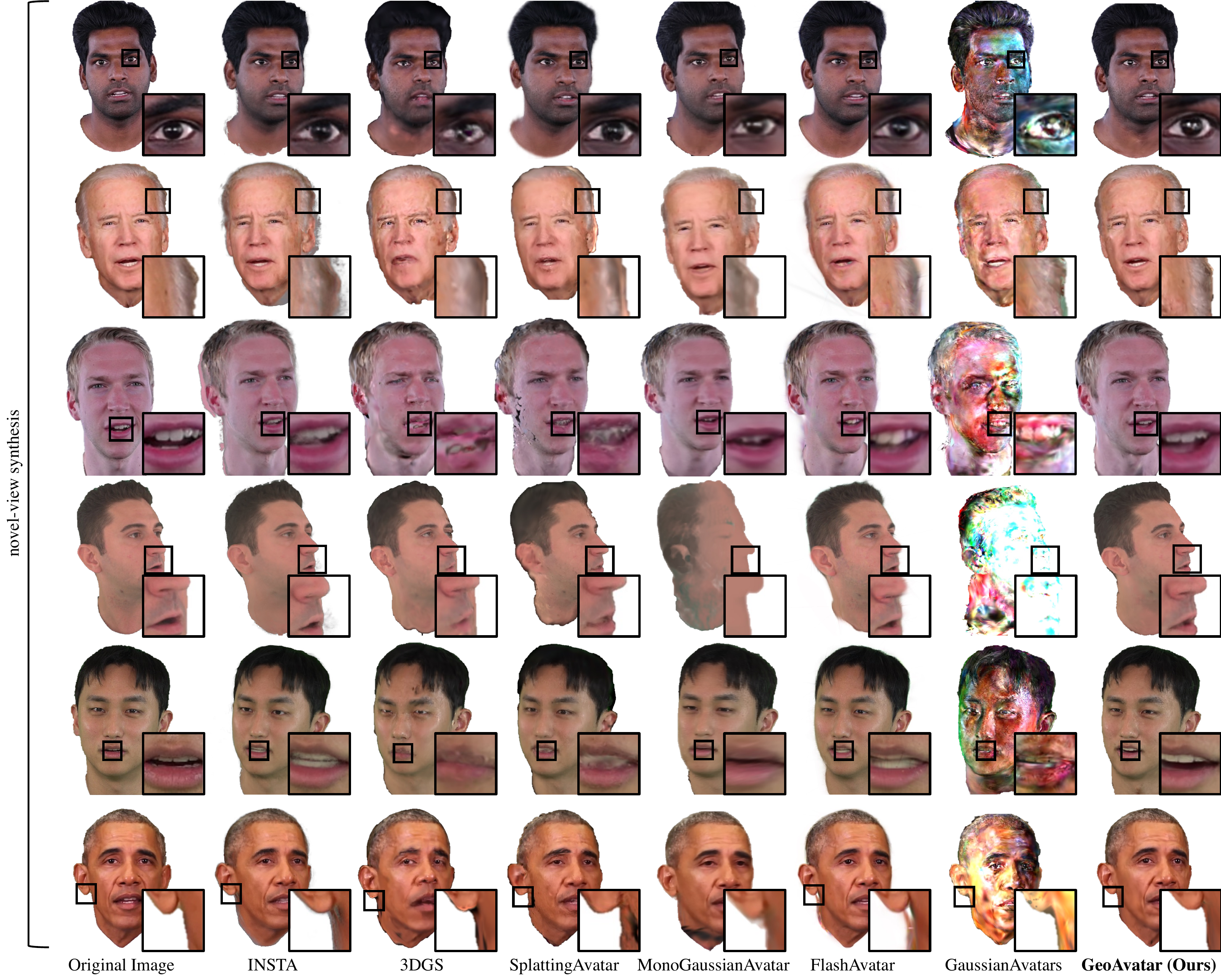}
\end{center}
\vspace{-0.5cm}
\caption{
\textbf{Additional novel-view synthesis results.}
We show additional novel-view synthesis results of ours, compared to baselines, on SplattingAvatar and DynamicFace datasets.
Ours shows clean and robust results not only on the facial region, \eg, the first, third, and fifth row, but also on the boundaries, \eg, the second, fourth, and sixth rows.
However, baselines suffer from either artifacts and low-resolution results in the facial region, or noisy boundaries in the boundary. Best viewed zoom-in.
}
\label{fig:appendix_novel}
\end{figure*}
\begin{figure*}[h]
\begin{center}
\includegraphics[width=0.9\textwidth]{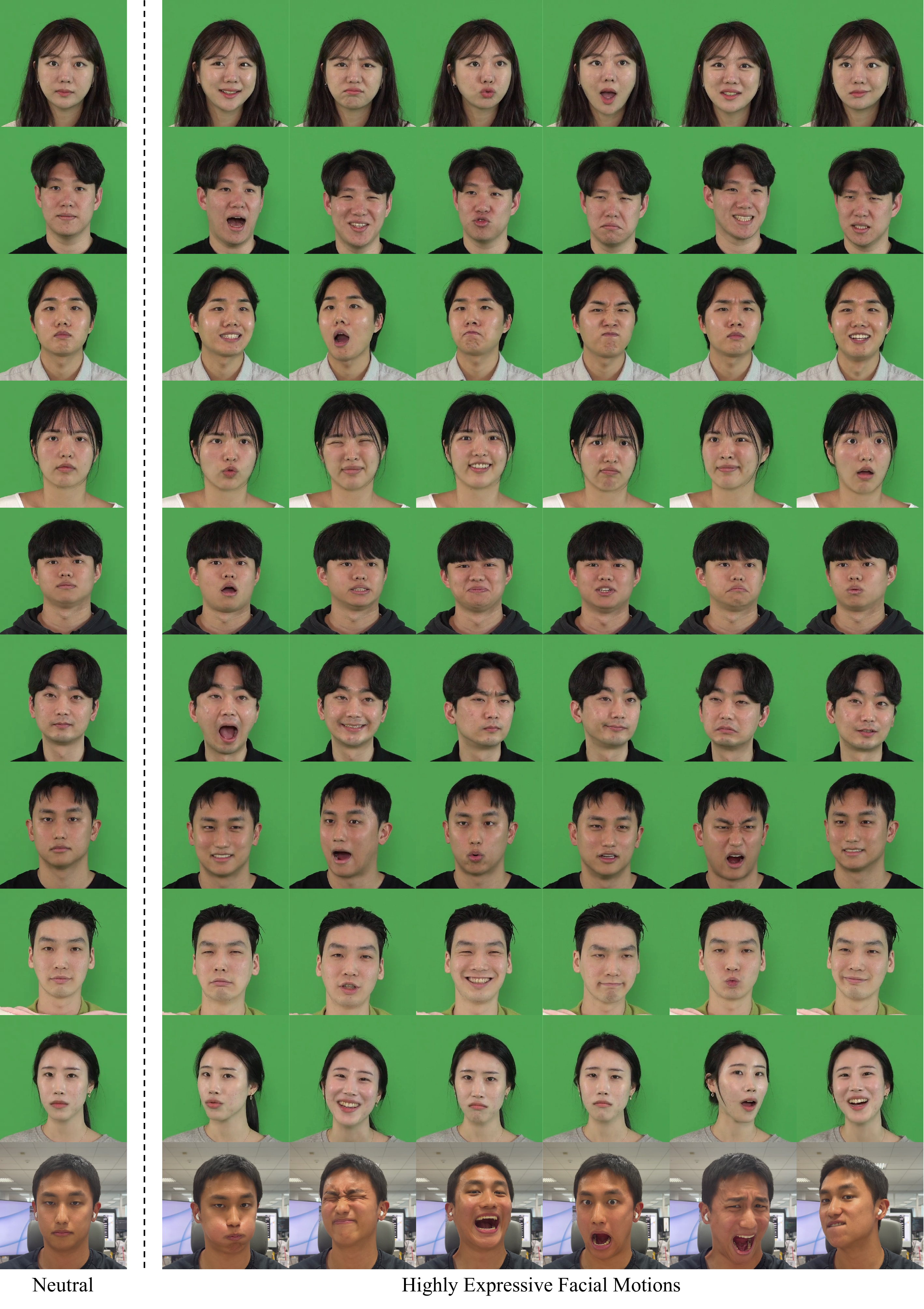}
\end{center}
\vspace{-0.5cm}
\caption{
\textbf{Examples of DynamicFace sequences.} We show our DynamicFace example sequences for all subjects. Our DynamicFace has diverse highly expressive facial motions.
} 
\label{fig:appendix_dynamicface}
\end{figure*}

\begin{figure*}[t]
\begin{center}
\includegraphics[width=1\textwidth]{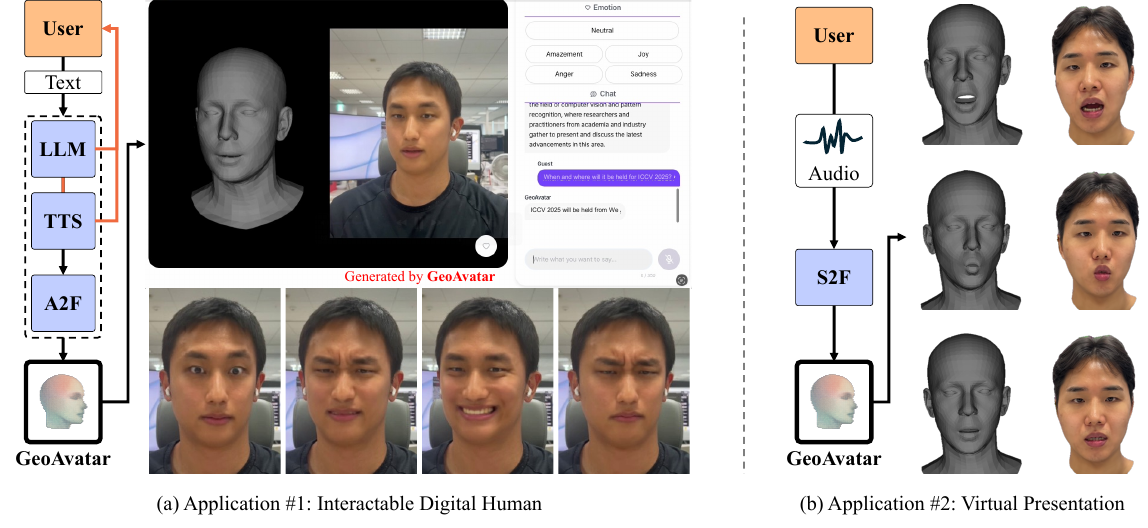}
\end{center}
\caption{
\textbf{Application examples.} Our proposed GeoAvatar framework shows its versatility across multiple applications. (a) Real-time Interactable Digital Human Demo: By integrating a large language model (LLM), text-to-speech (TTS), and NVIDIA Audio2Face (A2F) module, GeoAvatar enables real-time, interactive conversations with a fully animated digital human. Additional post-processing for backgrounds is used to enhance visual outputs. (b) Virtual Presentation: Given an input audio, GeoAvatar utilizes a speech-driven 3D facial animation module (S2F) to generate a high-quality digital human capable of delivering presentations with natural facial expressions and lip synchronization.
Both demo videos are visualized in our submitted project page HTML file.}
\label{fig:application}
\end{figure*}

\end{document}